\documentclass[aps,prb,twocolumn,superscriptaddress,notitlepage]{revtex4-1}
\usepackage{graphicx}
\usepackage[caption=false]{subfig}
\usepackage{float}
\usepackage{natbib}
\usepackage{xcolor}
\usepackage{xr}
\usepackage{amsmath}
\usepackage{amssymb}
\usepackage{array}
\usepackage{wasysym}
\usepackage{color,soul}
\usepackage{braket}
\usepackage{verbatim}
\usepackage{hyperref}
\usepackage{gensymb}
\usepackage{url}
\usepackage{dirtytalk}

\usepackage{filecontents}
\begin{document}

\title{Distinct Uniaxial Stress and Pressure Fingerprint of Superconductivity in the 3D Kagome Lattice Compound CeRu$_{2}$}

\author{O. Gerguri}
\thanks{These authors contributed equally to the experiments.}
\affiliation{PSI Center for Neutron and Muon Sciences CNM, 5232 Villigen PSI, Switzerland}

\author{D. Das}
\affiliation{PSI Center for Neutron and Muon Sciences CNM, 5232 Villigen PSI, Switzerland}

\author{V. Sazgari}
\thanks{These authors contributed equally to the experiments.}
\affiliation{PSI Center for Neutron and Muon Sciences CNM, 5232 Villigen PSI, Switzerland}

\author{H.X. Liu}
\affiliation{Beijing National Laboratory for Condensed Matter Physics and Institute of Physics, Chinese Academy of Sciences, Beijing 100190, China}
\affiliation{University of Chinese Academy of Sciences, Beijing 100049, China}

\author{C. Mielke III}
\affiliation{PSI Center for Neutron and Muon Sciences CNM, 5232 Villigen PSI, Switzerland}

\author{P. Král}
\affiliation{PSI Center for Neutron and Muon Sciences CNM, 5232 Villigen PSI, Switzerland}

\author{S.S. Islam}
\affiliation{PSI Center for Neutron and Muon Sciences CNM, 5232 Villigen PSI, Switzerland}

\author{J.N. Graham}
\affiliation{PSI Center for Neutron and Muon Sciences CNM, 5232 Villigen PSI, Switzerland}

\author{V.~Grinenko}
\affiliation{Tsung-Dao Lee Institute and School of Physics and Astronomy, Shanghai Jiao Tong University, Shanghai 201210, China}

\author{R.~Sarkar}
\affiliation{Institute for Solid State and Materials Physics, Technische Universität Dresden, D-01069 Dresden, Germany}

\author{T. Shiroka}
\affiliation{PSI Center for Neutron and Muon Sciences CNM, 5232 Villigen PSI, Switzerland}

\author{J.-X.~Yin}
\affiliation{Department of Physics, Southern University of Science and Technology, Shenzhen, Guangdong, 518055, China}

\author{J.~Chang}
\affiliation{Physik-Institut, Universit\"{a}t Z\"{u}rich, Winterthurerstrasse 190, CH-8057 Z\"{u}rich, Switzerland}

\author{R. Thomale}
\affiliation{Institut fur Theoretische Physik und Astrophysik, Universitat Wurzburg, Wurzburg, Germany}
\affiliation{Department of Physics, Indian Institute of Technology Madras, Chennai, India}

\author{H.H.~Klauss}
\affiliation{Institute for Solid State and Materials Physics, Technische Universität Dresden, D-01069 Dresden, Germany}

\author{R. Khasanov}
\affiliation{PSI Center for Neutron and Muon Sciences CNM, 5232 Villigen PSI, Switzerland}

\author{Y. Shi}
\email{ygshi@iphy.ac.cn}
\affiliation{Beijing National Laboratory for Condensed Matter Physics and Institute of Physics, Chinese Academy of Sciences, Beijing 100190, China}
\affiliation{University of Chinese Academy of Sciences, Beijing 100049, China}

\author{H. Luetkens}
\affiliation{PSI Center for Neutron and Muon Sciences CNM, 5232 Villigen PSI, Switzerland}

\author{Z. Guguchia}
\email{zurab.guguchia@psi.ch}
\affiliation{PSI Center for Neutron and Muon Sciences CNM, 5232 Villigen PSI, Switzerland}

\date{\today}

\begin{abstract}

The exploration of tunable superconductivity in strongly correlated electron systems is a central pursuit in condensed matter physics, with implications for both fundamental understanding and potential applications. The Laves phase CeRu$_{2}$, a pyrochlore compound, exhibits a three-dimensional (3D) Kagome lattice type geometry giving rise to flat bands and degenerate Dirac points, where band structure features intertwine with strong multi-orbital interaction effects deriving from its correlated electronic structure. Here, we combine muon spin rotation ($\mu$SR), uniaxial in-plane stress, and hydrostatic pressure to probe the superconducting state of CeRu$_{2}$. Uniaxial stress up to 0.22 GPa induces a dome-shaped evolution of the critical temperature $T_{\rm c}$, with an initial plateau, successively followed by enhancement and suppression without any structural phase transition. Stress is further found to drive a crossover from anisotropic to isotropic $s$-wave pairing. In contrast, hydrostatic pressure up to 2.2 GPa leaves $T_{\rm c}$ largely unchanged but alters the superfluid density from exponential to linear behavior at low temperatures, indicative of nodal superconductivity under hydrostatic pressure. Taken together, these results indicate that CeRu$_{2}$ occupies an ideal position in parameter space, enabling highly responsive and multifold tunability of superconductivity in this three-dimensional correlated electronic system. This warrants further quantitative analysis of the interplay between lattice geometry, electronic correlations, and pairing symmetry.

\end{abstract}
\maketitle
\section{Introduction}

The Kagome network, composed of corner-sharing triangles, naturally gives rise to a unique diversity of electronic features such as flat bands, van Hove singularities, and Dirac fermions, giving rise to peculiar interaction profiles resulting from destructive interference for the flat band and the associated sublattice frustration mechanism around the van Hove singularities \cite{kiesel2012sublattice}. In Kagome metals such as CsV$_{3}$Sb$_{5}$ \cite{ortiz2020cs} where several van Hove singularities are placed near the Fermi level, these effects conspire to reveal a wide range of unconventional electronic ground states including magnetism, nematic order, charge ordering, superconductivity, and more. In the uprising context of geometry-induced quantum effects, flat bands have become another central theme in the study of correlated electron systems, can be provided for in a Kagome-type lattice environment, opening up yet another realm of exotic phenomena \cite{guguchia2023tunable,mielke2021nodeless,guguchia2023unconventional,neupert2022charge,ortiz2020cs,kiesel2012sublattice}.


Although flat bands are typically associated with two-dimensional (2D) Kagome systems \cite{wang2023quantum,mielke2021nodeless,guguchia2023tunable,jovanovic2022simple,mielke2022time}—e.g., CoSn, FeSn, LaRu$_{3}$Si$_{2}$,  AV$_{3}$Sb$_{5}$ and AM$_6$X$_6$, recent attention has turned to three-dimensional analogs.\\

\indent In particular, the pyrochlore lattice which consists of corner-sharing tetrahedra and hosts a 3D-Kagome sublattice. They mimic the electronic isolation of 2D layers thanks to its unique crystallographic symmetry (space group Fd$\overline{3}$m) \cite{huang2024observation} and the geometrically driven destructive interference that suppresses inter-sublattice coupling\cite{kiesel2012sublattice}. Known pyrochlore compounds \cite{chudo2004geometric,wakefield2023three} such as CuV$_{2}$S$_{4}$ and CaNi$_{2}$ have also emerged as fertile ground for flat-band physics and topological states. In CaNi$_{2}$, flat bands and Dirac cones have been observed\cite{wakefield2023three} by angular resolved photoemission (ARPES) experiments.\\

\begin{figure*}[!]
    \centering
    \includegraphics[width=1.0\linewidth]{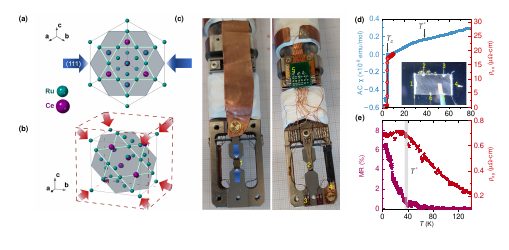}
    \vspace{-0.5cm}
    \caption{(a,b) The CeRu$_{2}$ unit cell with marked (111) Kagome planes (light grey), schematically showing the effect of uniaxial stress compression (a, stress applied in the (111) plane, although the exact direction within the plane is not known), and (b) the hydrostatic pressure. (c) Uniaxial stress device designed for the $\mu$SR experiments (1 - single-crystalline sample fixed to the detachable holder; 2 - coil for simultaneous measurements of AC magnetic susceptibility; 3 - force sensors; 4 - temperature sensor; 5 - printed circuit board). (d) Temperature dependence of AC susceptibility (left) and the longitudinal resistivity (right). Arrows mark the midpoint of the SC transition at $T_{\rm{c}}\approx$~4.86~K and the normal state anomaly at $T^{\rm{*}}\approx$~40~K. The inset shows the experimental setup for measurement of magnetotransport properties in CeRu$_{2}$ single crystal (1,4 - current leads; 2,3 - voltage probes for longitudinal resistivity, $\rho_{xx}$, measurements; 3,5 - voltage probes for Hall resistivity, $\rho_{xy}$). (e) Temperature dependence of magnetoresistance (MR, left) and Hall resistivity ($\rho_{xy}$, right) measured under the applied field of 9~T. The vertical grey line marks the $T^{\rm *}$ anomaly, corresponding to the onset of MR.}
    \label{fig:enter-label}
\end{figure*}

\indent In this perspective, the pyrochlore compound CeRu$_{2}$ \cite{matthias1958ferromagnetic}, has recently been re-examined due to its rich band structure arising from the Ru-based 3D Kagome network \cite{mielke2022local,guguchia2023unconventional}. Angle-resolved photoemission spectroscopy (ARPES) measurements along the (111) direction have revealed nearly dispersionless flat bands originating from both $3d$  Ru and $4f$  Ce states \cite{huang2024observation}, thus demonstrating the effect of the sublattice interference mechanism in preserving the 2D-Kagome flat band features onto the third dimension in the pyrochlore system. In addition, this structural framework not only supports flat bands, but also protects topological features, such as 3D Dirac cones at symmetry-enforced points in the Brillouin zone. In particular, CeRu$_{2}$ exhibits Dirac crossings constrained to the X point due to its non-symmorphic space group symmetry \cite{young2012dirac,saslow1966band}, a property shared between pyrochlore materials.

Superconductivity remains relatively rare in these systems and often competes with other symmetry-breaking orders. However, in CeRu$_{2}$, the normal-state electronic structure—dominated by hybridized $3d$ Ru and itinerant $4f$ Ce states—appears particularly favorable for the emergence of superconductivity. This is supported by complementary photoemission and de Haas–van Alphen (dHvA) measurements, which highlight the strong hybridization and itinerant character of the Ce-derived bands \cite{sekiyama2000probing,kang1999photoemission,inada1999haas}.
Additionally, anisotropic $s$-wave superconductivity in CeRu$_{2}$ has been reported by ${\mu}$SR, NMR/NQR, angle-resolved specific heat, and photoemission studies \cite{ishida1996ru,kiss2005photoemission,kittaka2013verification,manago2015effect}. The superfluid density is consistent with values observed in unconventional superconductors, suggesting an unconventional pairing mechanism. Furthermore, three distinct magnetic anomalies observed up to 8 T \cite{mielke2022local} indicate the possible presence of coexisting phases.
Taken together, these findings demonstrate that pyrochlore compounds with 3D Kagome sublattice represent a promising frontier for the study of flat-band physics, electronic correlation, and emergent superconductivity in three dimensions.

An important question in this study is whether these flat bands can be effectively tuned and, if so, how such tuning influences the microscopic properties of superconductivity. In this context, we present results from a unique combination of uniaxial stress applied along the Kagome plane and hydrostatic pressure ${\mu}$SR experiments (Fig. 1a-c), offering valuable insights into the tunable microscopic superconducting properties of CeRu$_{2}$. Key findings include (i) a non-monotonic dependence of the superconducting critical temperature on stress, (ii) an uniaxial stress-induced crossover from anisotropic to isotropic $s$-wave superconductivity, and (iii) a hydrostatic pressure-induced transition from nodeless to nodal unconventional pairing. Notably, the stress and pressure required to induce these changes differ by an order of magnitude, highlighting the distinct roles of directional and isotropic tuning parameters.

\begin{figure*}[!]
    \centering
    \includegraphics[width=1\linewidth]{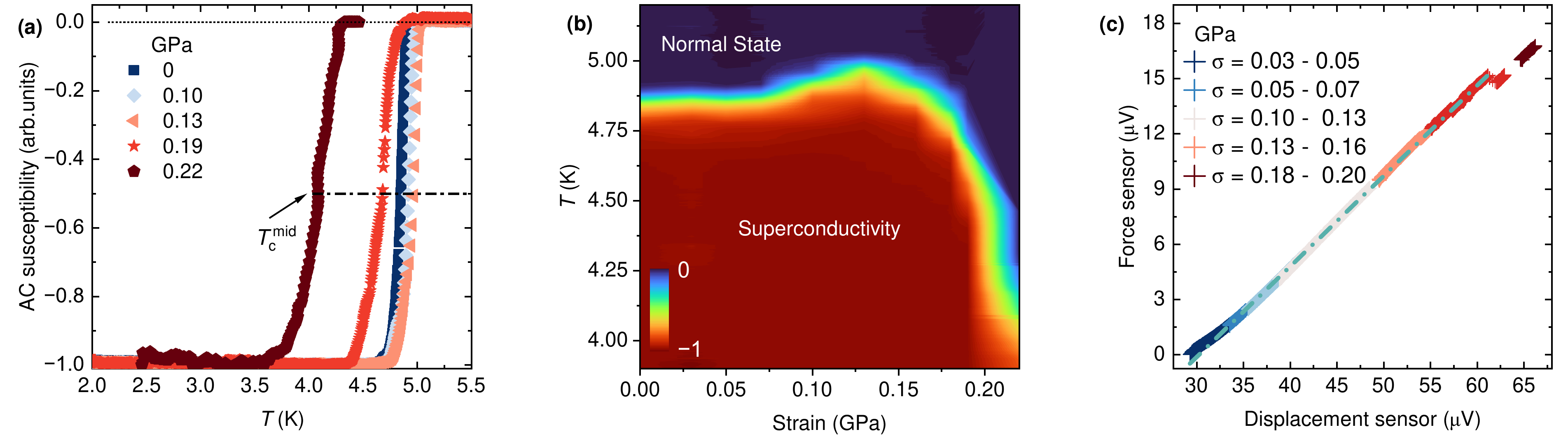}
    \caption{(a) The temperature dependence of AC susceptibility, measured under various uniaxial stress values. The dark dash-dotted line midway through the transition defines the estimation of $T_{\rm c}$. (b) Topographic color plot of normalized susceptibility of CeRu$_{2}$ as a function of temperature and applied stress. The light-color border corresponds to the transition from the superconducting to normal state and vice-versa. (c) Force-displacement curve for CeRu$_{2}$ measured from force and displacements sensors, where the latter is proportional to stress. The dashed line highlights the linear fit to the data across the applied stress region.}
    \label{fig:enter-label}
\end{figure*}

\section{Results and discussion}
\subsection{Uniaxial Stress}

The single crystal of CeRu$_{2}$ studied here exhibits a superconducting transition onset at $T_{\rm c}\approx$ 5~K, consistent with previous reports of high-quality samples. Samples with Ru deficiency typically show\cite{deng2024effect,sugawara1995superconducting} a higher $T_{\rm c}$  (6~K). Our sample also shows a clear anomaly in susceptibility around $T^{\rm*}=$ 40 K (see Figure 1d). Furthermore, Fig. 1e reveals that the magnetoresistance (MR) at 9 T increases below $T^{{*}}$, reaching 7$\%$ MR at 5 K. Superimposed on the plot, we show that the Hall resistivity that saturates under $T^{{*}}$ coincides with both the MR increase and the AC broad anomaly (see Figure 1e). Previous muon spin rotation measurements ($\mu$SR) \cite{huxley1997magnetic,mielke2022local} revealed a time-reversal symmetry (TRS) breaking in the same temperature regime. The combined results suggest a magnetic/electronic nature of the transition in the normal state pertaining this material. In the following, we focus on the stress dependence of the superconducting properties of CeRu$_{2}$.

Figure 2a presents the temperature dependence of magnetic susceptibility as a function of uniaxial strain applied within the Kagome (111) direction. The superconducting transition temperature, $T_{\rm c}$, is defined as the midpoint of the transition, corresponding to the exclusion of 50${\%}$ magnetic flux. As shown, $T_{\rm c}$ remains nearly constant up to 0.07 GPa, and then increases slightly, reaching a maximum at 0.13 GPa, which is then followed by a continuous decrease at higher strain. At 0.22 GPa, $T_{\rm c}$ is suppressed to 4~K, an approximate 16${\%}$ reduction compared to ambient pressure. This behavior reflects a dome-shaped stress dependence of $T_{\rm c}$. Figure 2c shows that the stress–strain relationship remains linear in the elastic regime of the sample, indicating the absence of any stress-induced structural phase transition in CeRu$_{2}$. Thus, the observed modulation of $T_{\rm c}$ is not related to structural changes but probably originates from stress-induced shifts in the electronic structure, particularly involving flat bands near the Fermi level.

In CeRu$_{2}$, the situation is further complicated by the presence of multiple flat-band features, as revealed by ARPES measurements\cite{huang2024observation}. Three distinct flat-band-like features are observed: (1) the Kondo resonance state ($4f_{5/2}^{1}$) located at the Fermi level, (2) its spin-orbit sideband ($4f_{7/2}^{1}$) at approximately $-0.25$ eV, both characteristic of Ce-based systems, and (3) a disentangled flat band originating from destructive interference in the Ru 3$d$ orbitals of the Kagome pyrochlore lattice, located at $-0.11$ eV. This proximity implies that a significant portion of the associated density of states (DOS) peak lies near the Fermi level. It is plausible that under the right stress conditions along the Kagome plane, this Ru-derived flat band and associated DOS peak could be shifted toward the Fermi level, thereby slightly enhancing the superconducting state through an increase in the density of states (DOS). However, beyond a critical stress threshold, the flat band may overshoot the Fermi level, shifting the DOS center of mass away and thereby suppressing $T_{\rm c}$. Whether this ~16${\%}$ reduction in $T_{\rm c}$ is driven solely by the movement of the Kagome flat band, or by collective changes involving both Ce and Ru flat-band features, remains an open question requiring theoretical investigation. A relevant comparison comes from CaNi$_2$, where a low-lying flat band at $-0.4$ eV was shifted to the Fermi level through only 2${\%}$ Ru doping in the isostructural compound\cite{wakefield2023three} Ca(Rh$_{0.98}$Ru$_{0.02}$)$_2$. This highlights the sensitivity of the electronic structure in such systems, which will be shown later that even a modest applied stress in CeRu$_2$ could significantly modify the Fermi surface topology by shifting multiple flat-band features.

\begin{figure*}[!]
    \centering
    \includegraphics[width=1\linewidth]{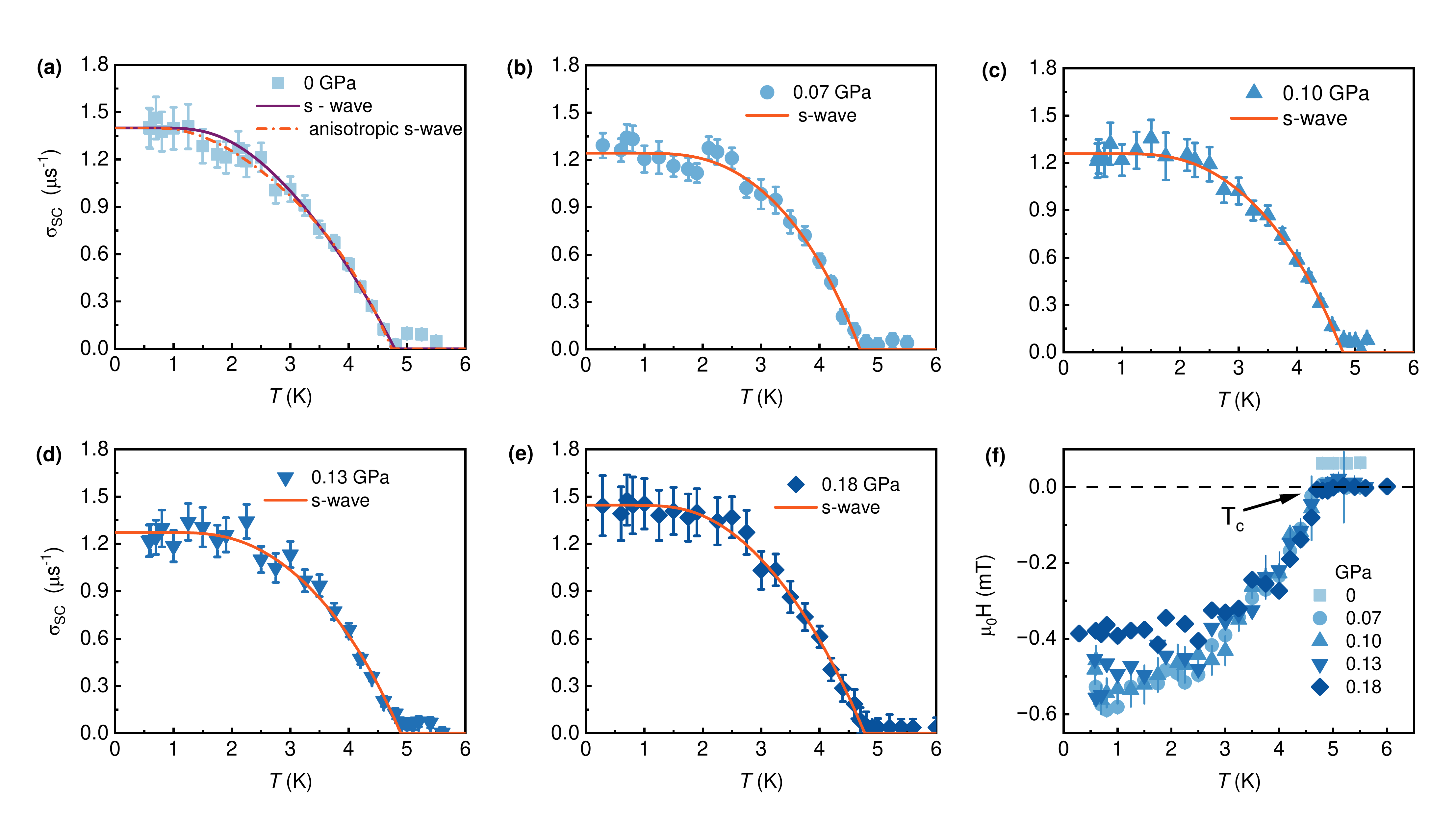}
    \caption{(a-e) The temperature dependence of the superconducting muon spin relaxation rate $\sigma_{\rm SC}$, recorded under various stress conditions up to 0.18~GPa. The dash-dotted and solid lines represent the fits using the anisotropic $s$-wave and isotropic $s$-wave models, respectively. (f) The temperature dependence of the diamagnetic shift, recorded under various uniaxial stress pressures.}
    \label{fig:enter-label}
\end{figure*}

Next, we examine how the superfluid density and superconducting gap structure evolve under uniaxial stress. This analysis is based on measurements of the superconducting muon spin relaxation rate in the vortex state of CeRu$_2$, using the transverse field TF-$\mu$SR. These time spectra were recorded in both the normal state (10 K) and the superconducting state (0.27 K), with a field of 30 mT applied along the stress direction. The normal-state spectrum shows weak depolarization due to the random local fields from the nuclear moments, and is well described by a single Gaussian function as shown in our previous study\cite{mielke2022local}. In contrast, the superconducting state exhibits a significantly enhanced relaxation rate due to the formation of the flux-line lattice (FLL) as reported previously\cite{mielke2022local}. The resulting asymmetric field distribution is modeled using a sum of two Gaussians, from which a single central field value is extracted, as detailed in the Methods section. The temperature dependence of the superconducting muon spin depolarization rate, $\sigma_{\rm SC}$, under five different stress conditions (up to 0.18 GPa) is presented in Figure 3. In order to investigate the symmetry of the SC gap, we underline the fact that the penetration depth $\lambda(T)$ is related to $\sigma_{\rm SC}(T)$ in the presence of an external applied field $H_{a}$, which induces a perfect triangular vortex lattice with field strength $H_{a}\ll H_{c2}$ by the equation\cite{brandt1988flux}:
\begin{equation}
\frac{\sigma_{\rm SC}}{\gamma_{\mu}} = 0.06091\frac{\Phi_{0}}{\lambda^{2}(T)}    
\end{equation}
where $\gamma_{\mu}$ is the gyromagnetic ratio of the muon and $\Phi_{0}$ is the magnetic-flux quantum. The temperature dependence of the superfluid density (related to $\lambda^{-2}(T)$) was then fitted with nodeless (an)isotropic $s$-wave and nodal $d$-wave models to determine the superconducting gap symmetry (see Methods section).\\

As shown in Table 1, the anisotropic $s$-wave model, which accounts for variation of the a single gap value from $\Delta_{\rm min}$ to $\Delta_{\rm max}$, provides the best fit to the data under zero stress, yielding a gap ratio of $\Delta_{\text{min}}/\Delta_{\text{max}} = 0.41$. This result is in very good agreement with previous $\mu$SR measurements without a uniaxial stress cell \cite{mielke2022local}, and is consistent with values reported from NMR ($\Delta_{\text{min}}/\Delta_{\text{max}} = 0.33$) \cite{kittaka2013verification} and photoemission experiments ($\Delta_{\text{min}}/\Delta_{\text{max}} = 0.447$) \cite{kiss2005photoemission}. In Figure 3a we highlight the difference between isotropic(solid line)/anisotropic s-wave(dashed line). Upon increasing the uniaxial stress within the (111)-planes, the superconducting gap evolves toward isotropy, as the anisotropy parameter $\Delta_{\text{min}}/\Delta_{\text{max}}$ vanishes at 0.07~GPa as seen in Table 1. This trend indicates a stress-induced crossover from an anisotropic to an isotropic $s$-wave superconducting state.

Further insight are gained from the temperature dependence of the diamagnetic shift, shown in the Figure 3f. The shift is defined as $\Delta B_{\text{dia}} = \mu_0(H_{\text{int,SC}} - H_{\text{int,NS}})$, where $H_{\text{int,SC}}$ and $H_{\text{int,NS}}$ are the internal magnetic fields in the superconducting and normal states, respectively. A pronounced diamagnetic response of approximately 0.6 mT develops below $T_{\text{c}}$, reflecting the onset of bulk superconductivity. Importantly, the magnitude and shape of this diamagnetic shift remain unaffected by applied stress.

Taken together, these results show that while uniaxial stress does not significantly alter the absolute value of the superfluid density or the diamagnetic response, it plays a crucial role in tuning the superconducting gap symmetry—driving a continuous transition from anisotropic to fully isotropic nodeless behavior.


\renewcommand{\arraystretch}{1.2}
\begin{table}
    \setlength{\tabcolsep}{2.4pt}
    \centering
    \begin{tabular}{c|c|c|c|}
    Symmetry & $\lambda_{0}$~(nm) & $\Delta_{max}$~(meV) & $\Delta_{min}$~(meV) \\
    \hline
    \hline
       \textit{s}-wave (0 Gpa) & 282(4) & 0.82(4) & - \\
      an.\textit{s}-wave (0 GPa)  & 278(6)  & 0.83(4) & 0.34(16)\\ \hline
         \textit{s}-wave (0.07 GPa)& 294(3)  & 0.95(5) & - \\
          an.\textit{s}-wave (0.07 GPa)& 294(3)  & 0.95(5) & - \\ \hline
           \textit{s}-wave (0.10 GPa)& 292(4) & 0.94(4) & - \\
            an.\textit{s}-wave (0.10 GPa)& 291(5) & 0.95(4) & - \\ \hline
            \textit{s}-wave (0.13 GPa)& 290(4) & 0.93(4) & - \\
            an.\textit{s}-wave (0.13 GPa)& 290(4) & 0.93(4) & - \\
            \hline
            \textit{s}-wave (0.18 GPa)& 272(5) & 0.85(5) & - \\
            an.\textit{s}-wave (0.18 GPa)& 272(5) & 0.86(5) & - \\
            \hline
    \end{tabular}
    \caption{Fit parameters of gap symmetries extracted from the temperature dependence of superconducting relaxation rate $\sigma_{\rm SC}$ under different stress conditions.}
    \label{tab:my_label}
\end{table}

 
 \begin{figure}[!]
    \centering \includegraphics[width = 0.9\linewidth]{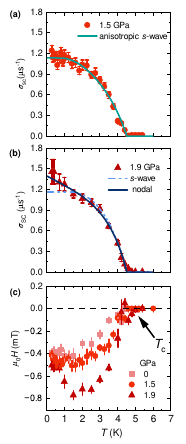}
    \caption{The temperature dependence of the superconducting muon spin relaxation rate $\sigma_{\rm SC}$, recorded under various hydrostatic pressure conditions: (a) 1.5~GPa and (b) 1.9~GPa. In panel (a), solid line corresponds to the fit using an anisotropic $s$-wave model. In panel (b), the dashed and the solid lines represent the fits using the isotropic $s$-wave and nodal models, respectively. (c) The temperature dependence of the diamagnetic shift, recorded under various hydrostatic pressures.}
    \label{fig:enter-label}
\end{figure}

\subsection{Hydrostatic pressure}
 Complementary to stress, hydrostatic pressure $\mu$SR \cite{khasanov2022three,khasanov2022perspective} experiment was conducted on CeRu$_{2}$. An external magnetic field of 30 mT was applied, and the sample was field-cooled from above $T_{\rm c}$ to establish a well-ordered vortex lattice. Measurements were carried out over the temperature range of 0.25–8 K. The Fourier transform of the muon asymmetry spectra reveals an asymmetric internal field distribution $f(B)$, arising from the flux-line lattice (FLL) in the superconducting state as previously reported\cite{mielke2021nodeless} and an additional background contribution from the pressure cell. Thus, for temperatures below $T_{\rm c}$, the transverse-field (TF) $\mu$SR spectra were analyzed using a three-component Gaussian model: two components corresponding to the superconducting vortex state and one accounting for the signal from the pressure cell. From this fit, the superconducting depolarization rate $\sigma_{\text{SC}}$ was extracted, allowing us to trace the evolution of the superconducting gap structure under hydrostatic pressure (Figure 4a).

In the analysis, the anisotropy ratio $\Delta_{\text{min}}/\Delta_{\text{max}}$ was fixed at 0.41 for 0 GPa, consistent with stress-dependent measurements and previous results under ambient conditions \cite{mielke2021nodeless}. Up to 1.5 GPa, the data are well described using an anisotropic $s$-wave model (Figure 4b). However, at the highest applied pressure of 1.9 GPa, the best fit is obtained with a nodal gap symmetry (Figure 4b). At this pressure, $\sigma_{\text{SC}}(T)$ exhibits a linear temperature dependence at low temperatures, indicating that $\lambda^{-2}(T)$ approaches its zero-temperature value linearly, a hallmark of nodes in the superconducting gap.
This pressure-induced nodal behavior suggests that hydrostatic pressure suppresses the minimum gap $\Delta_{\text{min}}$, eventually closing it and allowing for gapless quasiparticle excitations. This, in turn, gives rise to the observed power-law behavior of the relaxation rate at low temperatures.

The diamagnetic response under hydrostatic pressure also differs significantly from that under uniaxial stress. As pressure increases the overall magnitude of the diamagnetic shift nearly doubles at 1.9 GPa. Most notably, a paramagnetic upturn emerges below approximately 1.5 K. Paramagnetic shifts of similar nature have been reported in cuprate and iron-based high-temperature superconductors \cite{sonier2004field, kadono2005field, fujita2004magnetic, khasanov2009superconductivity} and was interpreted as field induced magnetism. Such anomalous paramagnetic responses have also been discussed as manifestations of unconventional pairing symmetry in the high-temperature superconductors \cite{sigrist1995unusual}.  In the absence of magnetism \cite{deng2024effect} in CeRu$_2$ within the investigated pressure range, the observed paramagnetic shift below 1.5 K, together with the emergence of nodal behavior, is likely associated with intrinsic modifications of the superconducting state.\\

\renewcommand{\arraystretch}{1.25}
\begin{table}[!]
    \setlength{\tabcolsep}{2.4pt}
    \centering
    \begin{tabular}{c|c|c|c|}
    Symmetry & $\lambda_{0}$~(nm) & $\Delta_{max}$~(meV) & $\Delta_{min}$~(meV) \\
    \hline
    \hline
       \textit{s}-wave (0 Gpa)& 300(4) & 0.72(2) & -\\
      an.\textit{s}-wave (0 GPa)  & 297(4) & 0.69(2) & 0.29(7)\\ \hline
         \textit{s}-wave (1.5 GPa)& 311(3) & 0.83(1) & - \\
          an.\textit{s}-wave (1.5 GPa)& 307(4) & 0.79(2) & 0.32(3)\\ \hline
            nodal (1.9 GPa)& 303(5) & 1.22(4) & 0\\
            \textit{s}-wave (1.9 GPa)& 277(4) & 0.92(3) & -\\
       \hline

    \end{tabular}
    \caption{Fit parameters of gap symmetries extracted from
the temperature dependence of superconducting relaxation
rate $\sigma_{\rm SC}$ at different pressures.}
    \label{tab:my_label}
\end{table}

A pressure-induced crossover from a nodeless to a nodal superconducting state has previously been observed\cite{guguchia2015direct} in the iron-based superconductor Ba$_{1-x}$Rb$_x$Fe$_2$As$_2$, where the appearance of accidental nodes was proposed, or as in the case of KFe$_{2}$As$_{2}$ where the non-monotonic pressure behavior of $T_{\rm c}$ is proposed\cite{tafti2013sudden} as an effect of the change of SC pairing symmetry. Additionally, in Kagome superconductors such as (K,Rb)V$_3$Sb$_5$, a pressure-induced transition from nodal to nodeless gap structure has been reported\cite{guguchia2023tunable} and attributed to the suppression of charge order. In our case, the emergence of nodal behavior under pressure does not necessarily imply a $d$-wave symmetry, as any nodal gap structure would produce the observed linear temperature dependence of the penetration depth. A detailed microscopic understanding of how pressure influences the superconducting state is still required. Altogether, these findings underscore the powerful role of hydrostatic pressure as a tuning parameter in correlated superconductors. In CeRu$_2$, this tuning appears to act as a modulator of the superconducting pairing symmetry.

\section{Conclusion}
We present the first comprehensive study of uniaxial stress and hydrostatic pressure effects on the microscopic superconducting properties of the pyrochlore compound CeRu$_{2}$, which hosts a three-dimensional (111) kagome lattice. Our key findings include: (i) a non-monotonic evolution of the superconducting transition temperature under uniaxial stress applied along the kagome plane; in the absence of any structural changes, this behavior suggests an unconventional superconducting state; (ii) stress-induced modifications of the superfluid density, consistent with a crossover from anisotropic $s$-wave to a more isotropic superconducting gap; and (iii) a pressure-induced change in the temperature dependence of the superfluid density at low temperatures, indicative of a transition to a nodal pairing state or the emergence of accidental nodes. These results reveal distinct and complementary pathways for tuning superconductivity through mechanical means, underscoring the need for theoretical insight into the interplay between lattice geometry, electronic correlations, and pairing symmetry in CeRu$_{2}$. In particular, a systematic ab initio analysis of the impact of point group symmetry breaking from uniaxial stress vs. the point group preserving hydrostatic pressure needs to focus on the itineracy vs. localized nature of electronic degress of freedom from a heavy fermion perspective, the modification of the correlated character via changing the interaction over bandwith ratio, and the band structure dynamics of flat bands, van Hove singularities, and eigenstate sublattice distribution both within the Brillouin zone and with respect to the Fermi level. Although similar phenomena are observed across various families of unconventional superconductors, CeRu$_{2}$ occupies a unique position at the intersection of heavy-fermion and kagome physics, which are both known for their rich and complex quantum states, combined with an enhanced of electronic order through the material’s 3D character. The mechanical tunability demonstrated here offers a powerful platform to explore unifying mechanisms underlying unconventional superconductivity in broader classes of materials, including cuprates, iron-based superconductors, and nickelates.

\section{Methods}

The single crystal of CeRu$_{2}$ was grown using the Czochralski pulling method with a prepared polycrystalline ingot i5n a tetra-arc furnace. Firstly, the spheroidal ingot of mass 10 g was prepared by arc melting stoichiometric amounts of lumpish cerium and ruthenium, which was then placed on the water-cooled copper hearth of the tetra-arc furnace under argon atmosphere with a titanium ingot used as an oxygen getter. The current of 16A was applied into the ingot to melt it and the velocity of the hearth rotation was 0.3 revolutions per minute. A columnar single crystal was successfully grown after several hours.\\

\textbf{Hydrostatic Pressure Cell:}
A double-wall piston-cylinder cell, constructed from CuBe and MP35N materials and specifically designed for ${\mu}$SR experiments under pressure, was utilized to generate pressures up to 1.9 GPa \cite{khasanov2022three,khasanov2022perspective}. Daphne 7373 served as the pressure-transmitting medium. The pressure calibration was established by measuring the superconducting transition of a small indium plate inside the sample volume from AC susceptibility. The pressure cell was optimized for maximum filling factor. Approximately 40${\%}$ of the muons stopped within the sample during the experiments.\\

\textbf{Uniaxial Stress Cell:}
For the ${\mu}$SR and AC susceptibility measurements, we used a cryo-compatible, piezoelectric-driven uniaxial pressure cell \cite{hicks2018piezoelectric,PhysRevLett.125.097005,grinenko2021split}, compatible with ${\mu}$SR experiments and designed to fit the He cryostat of the Dolly instrument at PSI’s ${\pi}$E1 beamline (see Figure 1c). The piezo actuators enable continuous in situ pressure tuning. The CeRu$_{2}$ single crystal (7 ${\times}$ 4.5 ${\times}$ 0.65 mm$^{3}$, oriented along [111]) was mounted on a detachable grade-2 titanium holder using Stycast-2850 FT epoxy. The holder, detailed in our previous work \cite{PhysRevLett.125.097005}, includes flexures to minimize torques and is designed to maximize muon acceptance by allowing missed muons to reach the veto counter. Hematite foils masked the sample plates to absorb muons outside the target area, resulting in 40${\%}$ muon stopping efficiency. A pair of 100-turn coils was positioned near the sample for in situ AC susceptibility measurements to track under various stress conditions. The muon-facing area was 4 ${\times}$ 4.5 mm$^{2}$.\\

\textbf{Magnetotransport measurements:}
Magnetotransport measurements were carried out in a standard four-probe method using the Physical Property Measurement System (PPMS, Quantum Design). The diagonal arrangement of voltage contacts was used and the magnetoresistance and Hall resistance were obtained by symmetrization and anti-symmetrization of the measured data, respectively.\\

\textbf{Muon Spin Rotation Experiments:}

Transverse-field (TF) $\mu$SR experiments with uniaxial strain experiment and high pressure were performed on the VMS instrument and the GPD instrument, equipped with an Oxford Sorption Pumped $^{3}$He Cryostat type Heliox/Variox, at the Swiss Muon Source (S$\mu$S) at the Paul Scherrer Institut, in Villigen, Switzerland.

In a ${\mu}$SR experiment nearly 100 ${\%}$ spin-polarized muons $\mu$$^{+}$ are implanted into the sample one at a time. The positively charged $\mu$$^{+}$ thermalize at interstitial lattice sites, where they act as magnetic microprobes. In a magnetic material the muon spin precesses in the local field $B_{\rm \mu}$ at the muon site with the Larmor frequency ${\nu}_{\rm \mu}$ = $\gamma_{\rm \mu}$/(2${\pi})$$B_{\rm \mu}$ (muon gyromagnetic ratio $\gamma_{\rm \mu}$/(2${\pi}$) = 135.5 MHz T$^{-1}$). Using the $\mu$SR technique, important length scales of superconductors can be measured, namely the magnetic penetration depth $\lambda$ and the coherence length $\xi$. If a type-II superconductor is cooled below $T_{\rm c}$ in an applied magnetic field ranged between the lower ($H_{c1}$) and the upper ($H_{c2}$) critical fields, a vortex lattice is formed which in general is incommensurate with the crystal lattice with vortex cores separated by much larger distances than those of the unit cell. Because the implanted muons stop at given crystallographic sites, they will randomly probe the field distribution of the vortex lattice. Such measurements need to be performed in a field applied perpendicular to the initial muon spin polarization (so called TF configuration).

Zero-field and high transverse-field experiments were carried out to probe normal state properties on the HAL instrument in a field range of 0.01 to 8 T. The sample in the form of compressed pellet (diameter 8 mm) was placed on the silver sample holder and mounted in the cryostat.\\

\textbf{Analysis of TF-${\mu}$SR data}:

In order to model the asymmetric field distribution ($P (B)$) in the SC state, the TF-${\mu}$SR time spectra measured below $T_{\rm c}$ are analyzed by using the following two-component functional form:

\begin{equation}
	\begin{aligned}
		A_{\rm TF} (t) = \sum_{i=0}^{2} A_{s,i}\exp\Big[-\frac{\sigma_{i}^2t^2}{2}\Big]\cos(\gamma_{\mu}B_{{\rm int},s,i}t+\varphi)  
		\label{eqS1}
	\end{aligned}
\end{equation}

Here $A_{s,i}$, $\sigma_{i}$ and $B_{{\rm int},s,i}$ is the initial asymmetry, relaxation rate, and local internal magnetic field of the $i$-th component. ${\varphi}$ is the initial phase of the muon-spin ensemble. $\gamma_{\mu}/(2{\pi})\simeq 135.5$~MHz/T is the muon gyromagnetic ratio. The first and second moments of the local magnetic field distribution are given by~\cite{Khasanov104504}

\begin{equation}
	\begin{aligned}
		\left\langle {B}\right\rangle = \sum_{i=0}^{2} \frac {A_{s,i}B_{{\rm int},s,i}}{A_{s,1}+A_{s,2}}
		\label{eqS2}
	\end{aligned}
\end{equation}
and 
\begin{equation}
	\begin{aligned}
		\left\langle {\Delta B}\right\rangle ^2 = \frac {\sigma ^2}{\gamma_{\mu}^2} =  \sum_{i=0}^2  \frac {A_{s,i}}{A_{s,1}+A_{s,2}} \Big[\sigma_i^2/\gamma_{\mu}^2 + \left(B_{{\rm int},s,i} - \langle B \rangle\right)^2\Big].
	\end{aligned}
\end{equation}

Above $T_{\rm c}$, in the normal state, the symmetric field distribution could be nicely modeled by only one component. The obtained relaxation rate and internal magnetic field are denoted by $\sigma_{\rm ns}$ and $B_{{\rm int},s,{\rm ns}}$. $\sigma_{\rm ns}$ is found to be small and temperature independent (dominated by nuclear magnetic moments) above $T_{\rm c}$ and assumed to be constant in the whole temperature range. Below $T_{\rm c}$, in the SC state, the relaxation rate and internal magnetic field are indicated by $\sigma_{\rm SC}$ and $B_{{\rm int},s,{\rm sc}}$. $\sigma_{\rm SC}$ is extracted by using $\sigma_\mathrm{SC} = \sqrt{\sigma^{2} - \sigma^{2}_\mathrm{ns}}$. $B_{{\rm int},s,{\rm sc}}$ is evaluated from $\left\langle {B}\right\rangle$ using Eq.~\eqref{eqS2}.\\

\textbf{Analysis of ${\lambda}(T)$}:

${\lambda}$($T$) was calculated within the local (London) approximation (${\lambda}$ ${\gg}$ ${\xi}$) by the following expression \cite{Suter69, Tinkham2004}:
\begin{equation}
	\frac{\sigma_{\rm SC}(T,\Delta_{0,i})}{\sigma_{\rm SC}(0,\Delta_{0,i})}=
	1+\frac{1}{\pi}\int_{0}^{2\pi}\int_{\Delta(_{T,\varphi})}^{\infty}(\frac{\partial f}{\partial E})\frac{EdEd\varphi}{\sqrt{E^2-\Delta_i(T,\varphi)^2}},
\end{equation}
where $f=[1+\exp(E/k_{\rm B}T)]^{-1}$ is the Fermi function, ${\varphi}$ is the angle along the Fermi surface, and ${\Delta}_{i}(T,{\varphi})={\Delta}_{0,i}{\Gamma}(T/T_{\rm c})g({\varphi}$)
(${\Delta}_{0,i}$ is the maximum gap value at $T=0$). The temperature dependence of the gap is approximated by the expression ${\Gamma}(T/T_{\rm c})=\tanh{\{}1.82[1.018(T_{\rm c}/T-1)]^{0.51}{\}}$,\cite{Carrington205} while $g({\varphi}$) describes the angular dependence of the gap and it is replaced by 1 for both an $s$-wave and an $s$+$s$-wave gap, and ${\mid}\cos(2{\varphi}){\mid}$ for a $d$-wave gap~\cite{guguchia2015direct}.\\

\section*{Acknowledgments}~
The ${\mu}$SR experiments were carried out at the Swiss Muon Source (S${\mu}$S) Paul Scherrer Insitute, Villigen, Switzerland. Z.G. acknowledges support from the Swiss National Science Foundation (SNSF) through SNSF Starting Grant (No. TMSGI2${\_}$211750). We acknowledge the usage of the instrumentation provided by the Electron Microscopy Facility at PSI and we thank the EMF team for their help and support with specimen cutting. Y.S. and H.X.L. acknowledge support from the National Key Research and Development Program of China ( No. 2024YFA140840), and the Synergetic Extreme Condition User Facility (SECUF, https://cstr.cn/31123.02.SECUF).\\

\section*{Author contributions}~
Z.G. conceived, designed, and supervised the project. Crystal growth: H.L. and Y.S. Magnetotransport experiments under ambient pressure: V.S., O.G. and Z.G. Preparation for uniaxial stress experiments: Z.G., D.D., and C.M.III. ${\mu}$SR and AC susceptibility experiments under uniaxial stress and corresponding discussions: Z.G., D.D., O.G., V.S., C.M.III., P.K., S.S.I., J.N.G., V.G., R.S., H.H.K., T.S., J.C., R.T., R.K., and H.L. ${\mu}$SR data analysis: O.G. and Z.G. Figure development and writing of the paper: O.G. and Z.G. with contributions from all authors. All authors discussed the results, interpretation, and conclusion.\\ 

\bibliography{References}{}

\begin{thebibliography}{43}%
\makeatletter
\providecommand \@ifxundefined [1]{%
 \@ifx{#1\undefined}
}%
\providecommand \@ifnum [1]{%
 \ifnum #1\expandafter \@firstoftwo
 \else \expandafter \@secondoftwo
 \fi
}%
\providecommand \@ifx [1]{%
 \ifx #1\expandafter \@firstoftwo
 \else \expandafter \@secondoftwo
 \fi
}%
\providecommand \natexlab [1]{#1}%
\providecommand \enquote  [1]{``#1''}%
\providecommand \bibnamefont  [1]{#1}%
\providecommand \bibfnamefont [1]{#1}%
\providecommand \citenamefont [1]{#1}%
\providecommand \href@noop [0]{\@secondoftwo}%
\providecommand \href [0]{\begingroup \@sanitize@url \@href}%
\providecommand \@href[1]{\@@startlink{#1}\@@href}%
\providecommand \@@href[1]{\endgroup#1\@@endlink}%
\providecommand \@sanitize@url [0]{\catcode `\\12\catcode `\$12\catcode `\&12\catcode `\#12\catcode `\^12\catcode `\_12\catcode `\%12\relax}%
\providecommand \@@startlink[1]{}%
\providecommand \@@endlink[0]{}%
\providecommand \url  [0]{\begingroup\@sanitize@url \@url }%
\providecommand \@url [1]{\endgroup\@href {#1}{\urlprefix }}%
\providecommand \urlprefix  [0]{URL }%
\providecommand \Eprint [0]{\href }%
\providecommand \doibase [0]{http://dx.doi.org/}%
\providecommand \selectlanguage [0]{\@gobble}%
\providecommand \bibinfo  [0]{\@secondoftwo}%
\providecommand \bibfield  [0]{\@secondoftwo}%
\providecommand \translation [1]{[#1]}%
\providecommand \BibitemOpen [0]{}%
\providecommand \bibitemStop [0]{}%
\providecommand \bibitemNoStop [0]{.\EOS\space}%
\providecommand \EOS [0]{\spacefactor3000\relax}%
\providecommand \BibitemShut  [1]{\csname bibitem#1\endcsname}%
\let\auto@bib@innerbib\@empty
\bibitem [{\citenamefont {Kiesel}\ and\ \citenamefont {Thomale}(2012)}]{kiesel2012sublattice}%
  \BibitemOpen
  \bibfield  {author} {\bibinfo {author} {\bibfnamefont {M.~L.}\ \bibnamefont {Kiesel}}\ and\ \bibinfo {author} {\bibfnamefont {R.}~\bibnamefont {Thomale}},\ }\href@noop {} {\bibfield  {journal} {\bibinfo  {journal} {Physical Review B—Condensed Matter and Materials Physics}\ }\textbf {\bibinfo {volume} {86}},\ \bibinfo {pages} {121105} (\bibinfo {year} {2012})}\BibitemShut {NoStop}%
\bibitem [{\citenamefont {Ortiz}\ \emph {et~al.}(2020)\citenamefont {Ortiz}, \citenamefont {Teicher}, \citenamefont {Hu}, \citenamefont {Zuo}, \citenamefont {Sarte}, \citenamefont {Schueller}, \citenamefont {Abeykoon}, \citenamefont {Krogstad}, \citenamefont {Rosenkranz}, \citenamefont {Osborn} \emph {et~al.}}]{ortiz2020cs}%
  \BibitemOpen
  \bibfield  {author} {\bibinfo {author} {\bibfnamefont {B.~R.}\ \bibnamefont {Ortiz}}, \bibinfo {author} {\bibfnamefont {S.~M.}\ \bibnamefont {Teicher}}, \bibinfo {author} {\bibfnamefont {Y.}~\bibnamefont {Hu}}, \bibinfo {author} {\bibfnamefont {J.~L.}\ \bibnamefont {Zuo}}, \bibinfo {author} {\bibfnamefont {P.~M.}\ \bibnamefont {Sarte}}, \bibinfo {author} {\bibfnamefont {E.~C.}\ \bibnamefont {Schueller}}, \bibinfo {author} {\bibfnamefont {A.~M.}\ \bibnamefont {Abeykoon}}, \bibinfo {author} {\bibfnamefont {M.~J.}\ \bibnamefont {Krogstad}}, \bibinfo {author} {\bibfnamefont {S.}~\bibnamefont {Rosenkranz}}, \bibinfo {author} {\bibfnamefont {R.}~\bibnamefont {Osborn}},  \emph {et~al.},\ }\href@noop {} {\bibfield  {journal} {\bibinfo  {journal} {Physical Review Letters}\ }\textbf {\bibinfo {volume} {125}},\ \bibinfo {pages} {247002} (\bibinfo {year} {2020})}\BibitemShut {NoStop}%
\bibitem [{\citenamefont {Guguchia}\ \emph {et~al.}(2023{\natexlab{a}})\citenamefont {Guguchia}, \citenamefont {Mielke~III}, \citenamefont {Das}, \citenamefont {Gupta}, \citenamefont {Yin}, \citenamefont {Liu}, \citenamefont {Yin}, \citenamefont {Christensen}, \citenamefont {Tu}, \citenamefont {Gong} \emph {et~al.}}]{guguchia2023tunable}%
  \BibitemOpen
  \bibfield  {author} {\bibinfo {author} {\bibfnamefont {Z.}~\bibnamefont {Guguchia}}, \bibinfo {author} {\bibfnamefont {C.}~\bibnamefont {Mielke~III}}, \bibinfo {author} {\bibfnamefont {D.}~\bibnamefont {Das}}, \bibinfo {author} {\bibfnamefont {R.}~\bibnamefont {Gupta}}, \bibinfo {author} {\bibfnamefont {J.-X.}\ \bibnamefont {Yin}}, \bibinfo {author} {\bibfnamefont {H.}~\bibnamefont {Liu}}, \bibinfo {author} {\bibfnamefont {Q.}~\bibnamefont {Yin}}, \bibinfo {author} {\bibfnamefont {M.~H.}\ \bibnamefont {Christensen}}, \bibinfo {author} {\bibfnamefont {Z.}~\bibnamefont {Tu}}, \bibinfo {author} {\bibfnamefont {C.}~\bibnamefont {Gong}},  \emph {et~al.},\ }\href@noop {} {\bibfield  {journal} {\bibinfo  {journal} {Nature communications}\ }\textbf {\bibinfo {volume} {14}},\ \bibinfo {pages} {153} (\bibinfo {year} {2023}{\natexlab{a}})}\BibitemShut {NoStop}%
\bibitem [{\citenamefont {Mielke~III}\ \emph {et~al.}(2021)\citenamefont {Mielke~III}, \citenamefont {Qin}, \citenamefont {Yin}, \citenamefont {Nakamura}, \citenamefont {Das}, \citenamefont {Guo}, \citenamefont {Khasanov}, \citenamefont {Chang}, \citenamefont {Wang}, \citenamefont {Jia} \emph {et~al.}}]{mielke2021nodeless}%
  \BibitemOpen
  \bibfield  {author} {\bibinfo {author} {\bibfnamefont {C.}~\bibnamefont {Mielke~III}}, \bibinfo {author} {\bibfnamefont {Y.}~\bibnamefont {Qin}}, \bibinfo {author} {\bibfnamefont {J.-X.}\ \bibnamefont {Yin}}, \bibinfo {author} {\bibfnamefont {H.}~\bibnamefont {Nakamura}}, \bibinfo {author} {\bibfnamefont {D.}~\bibnamefont {Das}}, \bibinfo {author} {\bibfnamefont {K.}~\bibnamefont {Guo}}, \bibinfo {author} {\bibfnamefont {R.}~\bibnamefont {Khasanov}}, \bibinfo {author} {\bibfnamefont {J.}~\bibnamefont {Chang}}, \bibinfo {author} {\bibfnamefont {Z.}~\bibnamefont {Wang}}, \bibinfo {author} {\bibfnamefont {S.}~\bibnamefont {Jia}},  \emph {et~al.},\ }\href@noop {} {\bibfield  {journal} {\bibinfo  {journal} {Physical Review Materials}\ }\textbf {\bibinfo {volume} {5}},\ \bibinfo {pages} {034803} (\bibinfo {year} {2021})}\BibitemShut {NoStop}%
\bibitem [{\citenamefont {Guguchia}\ \emph {et~al.}(2023{\natexlab{b}})\citenamefont {Guguchia}, \citenamefont {Khasanov},\ and\ \citenamefont {Luetkens}}]{guguchia2023unconventional}%
  \BibitemOpen
  \bibfield  {author} {\bibinfo {author} {\bibfnamefont {Z.}~\bibnamefont {Guguchia}}, \bibinfo {author} {\bibfnamefont {R.}~\bibnamefont {Khasanov}}, \ and\ \bibinfo {author} {\bibfnamefont {H.}~\bibnamefont {Luetkens}},\ }\href@noop {} {\bibfield  {journal} {\bibinfo  {journal} {npj Quantum Materials}\ }\textbf {\bibinfo {volume} {8}},\ \bibinfo {pages} {41} (\bibinfo {year} {2023}{\natexlab{b}})}\BibitemShut {NoStop}%
\bibitem [{\citenamefont {Neupert}\ \emph {et~al.}(2022)\citenamefont {Neupert}, \citenamefont {Denner}, \citenamefont {Yin}, \citenamefont {Thomale},\ and\ \citenamefont {Hasan}}]{neupert2022charge}%
  \BibitemOpen
  \bibfield  {author} {\bibinfo {author} {\bibfnamefont {T.}~\bibnamefont {Neupert}}, \bibinfo {author} {\bibfnamefont {M.~M.}\ \bibnamefont {Denner}}, \bibinfo {author} {\bibfnamefont {J.-X.}\ \bibnamefont {Yin}}, \bibinfo {author} {\bibfnamefont {R.}~\bibnamefont {Thomale}}, \ and\ \bibinfo {author} {\bibfnamefont {M.~Z.}\ \bibnamefont {Hasan}},\ }\href@noop {} {\bibfield  {journal} {\bibinfo  {journal} {Nature Physics}\ }\textbf {\bibinfo {volume} {18}},\ \bibinfo {pages} {137} (\bibinfo {year} {2022})}\BibitemShut {NoStop}%
\bibitem [{\citenamefont {Wang}\ \emph {et~al.}(2023)\citenamefont {Wang}, \citenamefont {Wu}, \citenamefont {McCandless}, \citenamefont {Chan},\ and\ \citenamefont {Ali}}]{wang2023quantum}%
  \BibitemOpen
  \bibfield  {author} {\bibinfo {author} {\bibfnamefont {Y.}~\bibnamefont {Wang}}, \bibinfo {author} {\bibfnamefont {H.}~\bibnamefont {Wu}}, \bibinfo {author} {\bibfnamefont {G.~T.}\ \bibnamefont {McCandless}}, \bibinfo {author} {\bibfnamefont {J.~Y.}\ \bibnamefont {Chan}}, \ and\ \bibinfo {author} {\bibfnamefont {M.~N.}\ \bibnamefont {Ali}},\ }\href@noop {} {\bibfield  {journal} {\bibinfo  {journal} {Nature Reviews Physics}\ }\textbf {\bibinfo {volume} {5}},\ \bibinfo {pages} {635} (\bibinfo {year} {2023})}\BibitemShut {NoStop}%
\bibitem [{\citenamefont {Jovanovic}\ and\ \citenamefont {Schoop}(2022)}]{jovanovic2022simple}%
  \BibitemOpen
  \bibfield  {author} {\bibinfo {author} {\bibfnamefont {M.}~\bibnamefont {Jovanovic}}\ and\ \bibinfo {author} {\bibfnamefont {L.~M.}\ \bibnamefont {Schoop}},\ }\href@noop {} {\bibfield  {journal} {\bibinfo  {journal} {Journal of the American Chemical Society}\ }\textbf {\bibinfo {volume} {144}},\ \bibinfo {pages} {10978} (\bibinfo {year} {2022})}\BibitemShut {NoStop}%
\bibitem [{\citenamefont {Mielke~III}\ \emph {et~al.}(2022{\natexlab{a}})\citenamefont {Mielke~III}, \citenamefont {Das}, \citenamefont {Yin}, \citenamefont {Liu}, \citenamefont {Gupta}, \citenamefont {Jiang}, \citenamefont {Medarde}, \citenamefont {Wu}, \citenamefont {Lei}, \citenamefont {Chang} \emph {et~al.}}]{mielke2022time}%
  \BibitemOpen
  \bibfield  {author} {\bibinfo {author} {\bibfnamefont {C.}~\bibnamefont {Mielke~III}}, \bibinfo {author} {\bibfnamefont {D.}~\bibnamefont {Das}}, \bibinfo {author} {\bibfnamefont {J.-X.}\ \bibnamefont {Yin}}, \bibinfo {author} {\bibfnamefont {H.}~\bibnamefont {Liu}}, \bibinfo {author} {\bibfnamefont {R.}~\bibnamefont {Gupta}}, \bibinfo {author} {\bibfnamefont {Y.-X.}\ \bibnamefont {Jiang}}, \bibinfo {author} {\bibfnamefont {M.}~\bibnamefont {Medarde}}, \bibinfo {author} {\bibfnamefont {X.}~\bibnamefont {Wu}}, \bibinfo {author} {\bibfnamefont {H.~C.}\ \bibnamefont {Lei}}, \bibinfo {author} {\bibfnamefont {J.}~\bibnamefont {Chang}},  \emph {et~al.},\ }\href@noop {} {\bibfield  {journal} {\bibinfo  {journal} {Nature}\ }\textbf {\bibinfo {volume} {602}},\ \bibinfo {pages} {245} (\bibinfo {year} {2022}{\natexlab{a}})}\BibitemShut {NoStop}%
\bibitem [{\citenamefont {Huang}\ \emph {et~al.}(2024)\citenamefont {Huang}, \citenamefont {Setty}, \citenamefont {Deng}, \citenamefont {You}, \citenamefont {Liu}, \citenamefont {Shao}, \citenamefont {Oh}, \citenamefont {Guo}, \citenamefont {Zhang}, \citenamefont {Yue} \emph {et~al.}}]{huang2024observation}%
  \BibitemOpen
  \bibfield  {author} {\bibinfo {author} {\bibfnamefont {J.}~\bibnamefont {Huang}}, \bibinfo {author} {\bibfnamefont {C.}~\bibnamefont {Setty}}, \bibinfo {author} {\bibfnamefont {L.}~\bibnamefont {Deng}}, \bibinfo {author} {\bibfnamefont {J.-Y.}\ \bibnamefont {You}}, \bibinfo {author} {\bibfnamefont {H.}~\bibnamefont {Liu}}, \bibinfo {author} {\bibfnamefont {S.}~\bibnamefont {Shao}}, \bibinfo {author} {\bibfnamefont {J.~S.}\ \bibnamefont {Oh}}, \bibinfo {author} {\bibfnamefont {Y.}~\bibnamefont {Guo}}, \bibinfo {author} {\bibfnamefont {Y.}~\bibnamefont {Zhang}}, \bibinfo {author} {\bibfnamefont {Z.}~\bibnamefont {Yue}},  \emph {et~al.},\ }\href@noop {} {\bibfield  {journal} {\bibinfo  {journal} {npj Quantum Materials}\ }\textbf {\bibinfo {volume} {9}},\ \bibinfo {pages} {71} (\bibinfo {year} {2024})}\BibitemShut {NoStop}%
\bibitem [{\citenamefont {Chudo}\ \emph {et~al.}(2004)\citenamefont {Chudo}, \citenamefont {Nakamura},\ and\ \citenamefont {Shiga}}]{chudo2004geometric}%
  \BibitemOpen
  \bibfield  {author} {\bibinfo {author} {\bibfnamefont {H.}~\bibnamefont {Chudo}}, \bibinfo {author} {\bibfnamefont {H.}~\bibnamefont {Nakamura}}, \ and\ \bibinfo {author} {\bibfnamefont {M.}~\bibnamefont {Shiga}},\ }\href@noop {} {\bibfield  {journal} {\bibinfo  {journal} {Solid state communications}\ }\textbf {\bibinfo {volume} {129}},\ \bibinfo {pages} {677} (\bibinfo {year} {2004})}\BibitemShut {NoStop}%
\bibitem [{\citenamefont {Wakefield}\ \emph {et~al.}(2023)\citenamefont {Wakefield}, \citenamefont {Kang}, \citenamefont {Neves}, \citenamefont {Oh}, \citenamefont {Fang}, \citenamefont {McTigue}, \citenamefont {Frank~Zhao}, \citenamefont {Lamichhane}, \citenamefont {Chen}, \citenamefont {Lee} \emph {et~al.}}]{wakefield2023three}%
  \BibitemOpen
  \bibfield  {author} {\bibinfo {author} {\bibfnamefont {J.~P.}\ \bibnamefont {Wakefield}}, \bibinfo {author} {\bibfnamefont {M.}~\bibnamefont {Kang}}, \bibinfo {author} {\bibfnamefont {P.~M.}\ \bibnamefont {Neves}}, \bibinfo {author} {\bibfnamefont {D.}~\bibnamefont {Oh}}, \bibinfo {author} {\bibfnamefont {S.}~\bibnamefont {Fang}}, \bibinfo {author} {\bibfnamefont {R.}~\bibnamefont {McTigue}}, \bibinfo {author} {\bibfnamefont {S.}~\bibnamefont {Frank~Zhao}}, \bibinfo {author} {\bibfnamefont {T.~N.}\ \bibnamefont {Lamichhane}}, \bibinfo {author} {\bibfnamefont {A.}~\bibnamefont {Chen}}, \bibinfo {author} {\bibfnamefont {S.}~\bibnamefont {Lee}},  \emph {et~al.},\ }\href@noop {} {\bibfield  {journal} {\bibinfo  {journal} {Nature}\ }\textbf {\bibinfo {volume} {623}},\ \bibinfo {pages} {301} (\bibinfo {year} {2023})}\BibitemShut {NoStop}%
\bibitem [{\citenamefont {Matthias}\ \emph {et~al.}(1958)\citenamefont {Matthias}, \citenamefont {Suhl},\ and\ \citenamefont {Corenzwit}}]{matthias1958ferromagnetic}%
  \BibitemOpen
  \bibfield  {author} {\bibinfo {author} {\bibfnamefont {B.}~\bibnamefont {Matthias}}, \bibinfo {author} {\bibfnamefont {H.}~\bibnamefont {Suhl}}, \ and\ \bibinfo {author} {\bibfnamefont {E.}~\bibnamefont {Corenzwit}},\ }\href@noop {} {\bibfield  {journal} {\bibinfo  {journal} {Physical Review Letters}\ }\textbf {\bibinfo {volume} {1}},\ \bibinfo {pages} {449} (\bibinfo {year} {1958})}\BibitemShut {NoStop}%
\bibitem [{\citenamefont {Mielke~III}\ \emph {et~al.}(2022{\natexlab{b}})\citenamefont {Mielke~III}, \citenamefont {Liu}, \citenamefont {Das}, \citenamefont {Yin}, \citenamefont {Deng}, \citenamefont {Spring}, \citenamefont {Gupta}, \citenamefont {Medarde}, \citenamefont {Chu}, \citenamefont {Khasanov} \emph {et~al.}}]{mielke2022local}%
  \BibitemOpen
  \bibfield  {author} {\bibinfo {author} {\bibfnamefont {C.}~\bibnamefont {Mielke~III}}, \bibinfo {author} {\bibfnamefont {H.}~\bibnamefont {Liu}}, \bibinfo {author} {\bibfnamefont {D.}~\bibnamefont {Das}}, \bibinfo {author} {\bibfnamefont {J.}~\bibnamefont {Yin}}, \bibinfo {author} {\bibfnamefont {L.}~\bibnamefont {Deng}}, \bibinfo {author} {\bibfnamefont {J.}~\bibnamefont {Spring}}, \bibinfo {author} {\bibfnamefont {R.}~\bibnamefont {Gupta}}, \bibinfo {author} {\bibfnamefont {M.}~\bibnamefont {Medarde}}, \bibinfo {author} {\bibfnamefont {C.}~\bibnamefont {Chu}}, \bibinfo {author} {\bibfnamefont {R.}~\bibnamefont {Khasanov}},  \emph {et~al.},\ }\href@noop {} {\bibfield  {journal} {\bibinfo  {journal} {Journal of Physics: Condensed Matter}\ }\textbf {\bibinfo {volume} {34}},\ \bibinfo {pages} {485601} (\bibinfo {year} {2022}{\natexlab{b}})}\BibitemShut {NoStop}%
\bibitem [{\citenamefont {Young}\ \emph {et~al.}(2012)\citenamefont {Young}, \citenamefont {Zaheer}, \citenamefont {Teo}, \citenamefont {Kane}, \citenamefont {Mele},\ and\ \citenamefont {Rappe}}]{young2012dirac}%
  \BibitemOpen
  \bibfield  {author} {\bibinfo {author} {\bibfnamefont {S.~M.}\ \bibnamefont {Young}}, \bibinfo {author} {\bibfnamefont {S.}~\bibnamefont {Zaheer}}, \bibinfo {author} {\bibfnamefont {J.~C.}\ \bibnamefont {Teo}}, \bibinfo {author} {\bibfnamefont {C.~L.}\ \bibnamefont {Kane}}, \bibinfo {author} {\bibfnamefont {E.~J.}\ \bibnamefont {Mele}}, \ and\ \bibinfo {author} {\bibfnamefont {A.~M.}\ \bibnamefont {Rappe}},\ }\href@noop {} {\bibfield  {journal} {\bibinfo  {journal} {Physical review letters}\ }\textbf {\bibinfo {volume} {108}},\ \bibinfo {pages} {140405} (\bibinfo {year} {2012})}\BibitemShut {NoStop}%
\bibitem [{\citenamefont {Saslow}\ \emph {et~al.}(1966)\citenamefont {Saslow}, \citenamefont {Bergstresser},\ and\ \citenamefont {Cohen}}]{saslow1966band}%
  \BibitemOpen
  \bibfield  {author} {\bibinfo {author} {\bibfnamefont {W.}~\bibnamefont {Saslow}}, \bibinfo {author} {\bibfnamefont {T.}~\bibnamefont {Bergstresser}}, \ and\ \bibinfo {author} {\bibfnamefont {M.~L.}\ \bibnamefont {Cohen}},\ }\href@noop {} {\bibfield  {journal} {\bibinfo  {journal} {Physical Review Letters}\ }\textbf {\bibinfo {volume} {16}},\ \bibinfo {pages} {354} (\bibinfo {year} {1966})}\BibitemShut {NoStop}%
\bibitem [{\citenamefont {Sekiyama}\ \emph {et~al.}(2000)\citenamefont {Sekiyama}, \citenamefont {Iwasaki}, \citenamefont {Matsuda}, \citenamefont {Saitoh}, \citenamefont {Onuki},\ and\ \citenamefont {Suga}}]{sekiyama2000probing}%
  \BibitemOpen
  \bibfield  {author} {\bibinfo {author} {\bibfnamefont {A.}~\bibnamefont {Sekiyama}}, \bibinfo {author} {\bibfnamefont {T.}~\bibnamefont {Iwasaki}}, \bibinfo {author} {\bibfnamefont {K.}~\bibnamefont {Matsuda}}, \bibinfo {author} {\bibfnamefont {Y.}~\bibnamefont {Saitoh}}, \bibinfo {author} {\bibfnamefont {Y.}~\bibnamefont {Onuki}}, \ and\ \bibinfo {author} {\bibfnamefont {S.}~\bibnamefont {Suga}},\ }\href@noop {} {\bibfield  {journal} {\bibinfo  {journal} {Nature}\ }\textbf {\bibinfo {volume} {403}},\ \bibinfo {pages} {396} (\bibinfo {year} {2000})}\BibitemShut {NoStop}%
\bibitem [{\citenamefont {Kang}\ \emph {et~al.}(1999)\citenamefont {Kang}, \citenamefont {Olson}, \citenamefont {Hedo}, \citenamefont {Inada}, \citenamefont {Yamamoto}, \citenamefont {Haga}, \citenamefont {{\=O}nuki}, \citenamefont {Kwon},\ and\ \citenamefont {Min}}]{kang1999photoemission}%
  \BibitemOpen
  \bibfield  {author} {\bibinfo {author} {\bibfnamefont {J.-S.}\ \bibnamefont {Kang}}, \bibinfo {author} {\bibfnamefont {C.}~\bibnamefont {Olson}}, \bibinfo {author} {\bibfnamefont {M.}~\bibnamefont {Hedo}}, \bibinfo {author} {\bibfnamefont {Y.}~\bibnamefont {Inada}}, \bibinfo {author} {\bibfnamefont {E.}~\bibnamefont {Yamamoto}}, \bibinfo {author} {\bibfnamefont {Y.}~\bibnamefont {Haga}}, \bibinfo {author} {\bibfnamefont {Y.}~\bibnamefont {{\=O}nuki}}, \bibinfo {author} {\bibfnamefont {S.}~\bibnamefont {Kwon}}, \ and\ \bibinfo {author} {\bibfnamefont {B.}~\bibnamefont {Min}},\ }\href@noop {} {\bibfield  {journal} {\bibinfo  {journal} {Physical Review B}\ }\textbf {\bibinfo {volume} {60}},\ \bibinfo {pages} {5348} (\bibinfo {year} {1999})}\BibitemShut {NoStop}%
\bibitem [{\citenamefont {Inada}\ and\ \citenamefont {{\=O}nuki}(1999)}]{inada1999haas}%
  \BibitemOpen
  \bibfield  {author} {\bibinfo {author} {\bibfnamefont {Y.}~\bibnamefont {Inada}}\ and\ \bibinfo {author} {\bibfnamefont {Y.}~\bibnamefont {{\=O}nuki}},\ }\href@noop {} {\bibfield  {journal} {\bibinfo  {journal} {Low Temperature Physics}\ }\textbf {\bibinfo {volume} {25}},\ \bibinfo {pages} {573} (\bibinfo {year} {1999})}\BibitemShut {NoStop}%
\bibitem [{\citenamefont {Ishida}\ \emph {et~al.}(1996)\citenamefont {Ishida}, \citenamefont {Mukuda}, \citenamefont {Kitaoka}, \citenamefont {Asayama},\ and\ \citenamefont {Onuki}}]{ishida1996ru}%
  \BibitemOpen
  \bibfield  {author} {\bibinfo {author} {\bibfnamefont {K.}~\bibnamefont {Ishida}}, \bibinfo {author} {\bibfnamefont {H.}~\bibnamefont {Mukuda}}, \bibinfo {author} {\bibfnamefont {Y.}~\bibnamefont {Kitaoka}}, \bibinfo {author} {\bibfnamefont {K.}~\bibnamefont {Asayama}}, \ and\ \bibinfo {author} {\bibfnamefont {Y.}~\bibnamefont {Onuki}},\ }\href@noop {} {\bibfield  {journal} {\bibinfo  {journal} {Zeitschrift f{\"u}r Naturforschung A}\ }\textbf {\bibinfo {volume} {51}},\ \bibinfo {pages} {793} (\bibinfo {year} {1996})}\BibitemShut {NoStop}%
\bibitem [{\citenamefont {Kiss}\ \emph {et~al.}(2005)\citenamefont {Kiss}, \citenamefont {Kanetaka}, \citenamefont {Yokoya}, \citenamefont {Shimojima}, \citenamefont {Kanai}, \citenamefont {Shin}, \citenamefont {Onuki}, \citenamefont {Togashi}, \citenamefont {Zhang}, \citenamefont {Chen} \emph {et~al.}}]{kiss2005photoemission}%
  \BibitemOpen
  \bibfield  {author} {\bibinfo {author} {\bibfnamefont {T.}~\bibnamefont {Kiss}}, \bibinfo {author} {\bibfnamefont {F.}~\bibnamefont {Kanetaka}}, \bibinfo {author} {\bibfnamefont {T.}~\bibnamefont {Yokoya}}, \bibinfo {author} {\bibfnamefont {T.}~\bibnamefont {Shimojima}}, \bibinfo {author} {\bibfnamefont {K.}~\bibnamefont {Kanai}}, \bibinfo {author} {\bibfnamefont {S.}~\bibnamefont {Shin}}, \bibinfo {author} {\bibfnamefont {Y.}~\bibnamefont {Onuki}}, \bibinfo {author} {\bibfnamefont {T.}~\bibnamefont {Togashi}}, \bibinfo {author} {\bibfnamefont {C.}~\bibnamefont {Zhang}}, \bibinfo {author} {\bibfnamefont {C.}~\bibnamefont {Chen}},  \emph {et~al.},\ }\href@noop {} {\bibfield  {journal} {\bibinfo  {journal} {Physical review letters}\ }\textbf {\bibinfo {volume} {94}},\ \bibinfo {pages} {057001} (\bibinfo {year} {2005})}\BibitemShut {NoStop}%
\bibitem [{\citenamefont {Kittaka}\ \emph {et~al.}(2013)\citenamefont {Kittaka}, \citenamefont {Sakakibara}, \citenamefont {Hedo}, \citenamefont {{\=O}nuki},\ and\ \citenamefont {Machida}}]{kittaka2013verification}%
  \BibitemOpen
  \bibfield  {author} {\bibinfo {author} {\bibfnamefont {S.}~\bibnamefont {Kittaka}}, \bibinfo {author} {\bibfnamefont {T.}~\bibnamefont {Sakakibara}}, \bibinfo {author} {\bibfnamefont {M.}~\bibnamefont {Hedo}}, \bibinfo {author} {\bibfnamefont {Y.}~\bibnamefont {{\=O}nuki}}, \ and\ \bibinfo {author} {\bibfnamefont {K.}~\bibnamefont {Machida}},\ }\href@noop {} {\bibfield  {journal} {\bibinfo  {journal} {Journal of the Physical Society of Japan}\ }\textbf {\bibinfo {volume} {82}},\ \bibinfo {pages} {123706} (\bibinfo {year} {2013})}\BibitemShut {NoStop}%
\bibitem [{\citenamefont {Manago}\ \emph {et~al.}(2015)\citenamefont {Manago}, \citenamefont {Ishida}, \citenamefont {Matsuda},\ and\ \citenamefont {{\=O}nuki}}]{manago2015effect}%
  \BibitemOpen
  \bibfield  {author} {\bibinfo {author} {\bibfnamefont {M.}~\bibnamefont {Manago}}, \bibinfo {author} {\bibfnamefont {K.}~\bibnamefont {Ishida}}, \bibinfo {author} {\bibfnamefont {T.~D.}\ \bibnamefont {Matsuda}}, \ and\ \bibinfo {author} {\bibfnamefont {Y.}~\bibnamefont {{\=O}nuki}},\ }\href@noop {} {\bibfield  {journal} {\bibinfo  {journal} {Journal of the Physical Society of Japan}\ }\textbf {\bibinfo {volume} {84}},\ \bibinfo {pages} {115001} (\bibinfo {year} {2015})}\BibitemShut {NoStop}%
\bibitem [{\citenamefont {Deng}\ \emph {et~al.}(2024)\citenamefont {Deng}, \citenamefont {Gooch}, \citenamefont {Liu}, \citenamefont {Salke}, \citenamefont {Bontke}, \citenamefont {Shao}, \citenamefont {You}, \citenamefont {Schulze}, \citenamefont {Kumar}, \citenamefont {Yin} \emph {et~al.}}]{deng2024effect}%
  \BibitemOpen
  \bibfield  {author} {\bibinfo {author} {\bibfnamefont {L.}~\bibnamefont {Deng}}, \bibinfo {author} {\bibfnamefont {M.}~\bibnamefont {Gooch}}, \bibinfo {author} {\bibfnamefont {H.}~\bibnamefont {Liu}}, \bibinfo {author} {\bibfnamefont {N.~P.}\ \bibnamefont {Salke}}, \bibinfo {author} {\bibfnamefont {T.}~\bibnamefont {Bontke}}, \bibinfo {author} {\bibfnamefont {S.}~\bibnamefont {Shao}}, \bibinfo {author} {\bibfnamefont {J.}~\bibnamefont {You}}, \bibinfo {author} {\bibfnamefont {D.~J.}\ \bibnamefont {Schulze}}, \bibinfo {author} {\bibfnamefont {R.}~\bibnamefont {Kumar}}, \bibinfo {author} {\bibfnamefont {J.-X.}\ \bibnamefont {Yin}},  \emph {et~al.},\ }\href@noop {} {\bibfield  {journal} {\bibinfo  {journal} {Materials Today Physics}\ }\textbf {\bibinfo {volume} {40}},\ \bibinfo {pages} {101322} (\bibinfo {year} {2024})}\BibitemShut {NoStop}%
\bibitem [{\citenamefont {Sugawara}\ \emph {et~al.}(1995)\citenamefont {Sugawara}, \citenamefont {Sato}, \citenamefont {Yamazaki}, \citenamefont {Kimura}, \citenamefont {Settai},\ and\ \citenamefont {{\=O}nuke}}]{sugawara1995superconducting}%
  \BibitemOpen
  \bibfield  {author} {\bibinfo {author} {\bibfnamefont {H.}~\bibnamefont {Sugawara}}, \bibinfo {author} {\bibfnamefont {H.}~\bibnamefont {Sato}}, \bibinfo {author} {\bibfnamefont {T.}~\bibnamefont {Yamazaki}}, \bibinfo {author} {\bibfnamefont {N.}~\bibnamefont {Kimura}}, \bibinfo {author} {\bibfnamefont {R.}~\bibnamefont {Settai}}, \ and\ \bibinfo {author} {\bibfnamefont {Y.}~\bibnamefont {{\=O}nuke}},\ }\href@noop {} {\bibfield  {journal} {\bibinfo  {journal} {Journal of the Physical Society of Japan}\ }\textbf {\bibinfo {volume} {64}},\ \bibinfo {pages} {4849} (\bibinfo {year} {1995})}\BibitemShut {NoStop}%
\bibitem [{\citenamefont {Huxley}\ \emph {et~al.}(1997)\citenamefont {Huxley}, \citenamefont {Boucherle}, \citenamefont {Bonnet}, \citenamefont {Bourdarot}, \citenamefont {Schustler}, \citenamefont {Caplan}, \citenamefont {Lelievre}, \citenamefont {Bernhoeft}, \citenamefont {Lejay},\ and\ \citenamefont {Gillon}}]{huxley1997magnetic}%
  \BibitemOpen
  \bibfield  {author} {\bibinfo {author} {\bibfnamefont {A.}~\bibnamefont {Huxley}}, \bibinfo {author} {\bibfnamefont {J.}~\bibnamefont {Boucherle}}, \bibinfo {author} {\bibfnamefont {M.}~\bibnamefont {Bonnet}}, \bibinfo {author} {\bibfnamefont {F.}~\bibnamefont {Bourdarot}}, \bibinfo {author} {\bibfnamefont {I.}~\bibnamefont {Schustler}}, \bibinfo {author} {\bibfnamefont {D.}~\bibnamefont {Caplan}}, \bibinfo {author} {\bibfnamefont {E.}~\bibnamefont {Lelievre}}, \bibinfo {author} {\bibfnamefont {N.}~\bibnamefont {Bernhoeft}}, \bibinfo {author} {\bibfnamefont {P.}~\bibnamefont {Lejay}}, \ and\ \bibinfo {author} {\bibfnamefont {B.}~\bibnamefont {Gillon}},\ }\href@noop {} {\bibfield  {journal} {\bibinfo  {journal} {Journal of Physics: Condensed Matter}\ }\textbf {\bibinfo {volume} {9}},\ \bibinfo {pages} {4185} (\bibinfo {year} {1997})}\BibitemShut {NoStop}%
\bibitem [{\citenamefont {Brandt}(1988)}]{brandt1988flux}%
  \BibitemOpen
  \bibfield  {author} {\bibinfo {author} {\bibfnamefont {E.}~\bibnamefont {Brandt}},\ }\href@noop {} {\bibfield  {journal} {\bibinfo  {journal} {Physical Review B}\ }\textbf {\bibinfo {volume} {37}},\ \bibinfo {pages} {2349} (\bibinfo {year} {1988})}\BibitemShut {NoStop}%
\bibitem [{\citenamefont {Khasanov}\ \emph {et~al.}(2022)\citenamefont {Khasanov}, \citenamefont {Urquhart}, \citenamefont {Elender},\ and\ \citenamefont {Kamenev}}]{khasanov2022three}%
  \BibitemOpen
  \bibfield  {author} {\bibinfo {author} {\bibfnamefont {R.}~\bibnamefont {Khasanov}}, \bibinfo {author} {\bibfnamefont {R.}~\bibnamefont {Urquhart}}, \bibinfo {author} {\bibfnamefont {M.}~\bibnamefont {Elender}}, \ and\ \bibinfo {author} {\bibfnamefont {K.}~\bibnamefont {Kamenev}},\ }\href@noop {} {\bibfield  {journal} {\bibinfo  {journal} {High Pressure Research}\ }\textbf {\bibinfo {volume} {42}},\ \bibinfo {pages} {29} (\bibinfo {year} {2022})}\BibitemShut {NoStop}%
\bibitem [{\citenamefont {Khasanov}(2022)}]{khasanov2022perspective}%
  \BibitemOpen
  \bibfield  {author} {\bibinfo {author} {\bibfnamefont {R.}~\bibnamefont {Khasanov}},\ }\href@noop {} {\bibfield  {journal} {\bibinfo  {journal} {Journal of Applied Physics}\ }\textbf {\bibinfo {volume} {132}} (\bibinfo {year} {2022})}\BibitemShut {NoStop}%
\bibitem [{\citenamefont {Sonier}\ \emph {et~al.}(2004)\citenamefont {Sonier}, \citenamefont {Poon}, \citenamefont {Luke}, \citenamefont {Kyriakou}, \citenamefont {Miller}, \citenamefont {Liang}, \citenamefont {Wiebe}, \citenamefont {Fournier},\ and\ \citenamefont {Greene}}]{sonier2004field}%
  \BibitemOpen
  \bibfield  {author} {\bibinfo {author} {\bibfnamefont {J.}~\bibnamefont {Sonier}}, \bibinfo {author} {\bibfnamefont {K.}~\bibnamefont {Poon}}, \bibinfo {author} {\bibfnamefont {G.}~\bibnamefont {Luke}}, \bibinfo {author} {\bibfnamefont {P.}~\bibnamefont {Kyriakou}}, \bibinfo {author} {\bibfnamefont {R.}~\bibnamefont {Miller}}, \bibinfo {author} {\bibfnamefont {R.}~\bibnamefont {Liang}}, \bibinfo {author} {\bibfnamefont {C.}~\bibnamefont {Wiebe}}, \bibinfo {author} {\bibfnamefont {P.}~\bibnamefont {Fournier}}, \ and\ \bibinfo {author} {\bibfnamefont {R.}~\bibnamefont {Greene}},\ }\href@noop {} {\bibfield  {journal} {\bibinfo  {journal} {Physica C: Superconductivity}\ }\textbf {\bibinfo {volume} {408}},\ \bibinfo {pages} {783} (\bibinfo {year} {2004})}\BibitemShut {NoStop}%
\bibitem [{\citenamefont {Kadono}\ \emph {et~al.}(2005)\citenamefont {Kadono}, \citenamefont {Ohishi}, \citenamefont {Koda}, \citenamefont {R.~Saha}, \citenamefont {Higemoto}, \citenamefont {Fujita},\ and\ \citenamefont {Yamada}}]{kadono2005field}%
  \BibitemOpen
  \bibfield  {author} {\bibinfo {author} {\bibfnamefont {R.}~\bibnamefont {Kadono}}, \bibinfo {author} {\bibfnamefont {K.}~\bibnamefont {Ohishi}}, \bibinfo {author} {\bibfnamefont {A.}~\bibnamefont {Koda}}, \bibinfo {author} {\bibfnamefont {S.}~\bibnamefont {R.~Saha}}, \bibinfo {author} {\bibfnamefont {W.}~\bibnamefont {Higemoto}}, \bibinfo {author} {\bibfnamefont {M.}~\bibnamefont {Fujita}}, \ and\ \bibinfo {author} {\bibfnamefont {K.}~\bibnamefont {Yamada}},\ }\href@noop {} {\bibfield  {journal} {\bibinfo  {journal} {Journal of the Physical Society of Japan}\ }\textbf {\bibinfo {volume} {74}},\ \bibinfo {pages} {2806} (\bibinfo {year} {2005})}\BibitemShut {NoStop}%
\bibitem [{\citenamefont {Fujita}\ \emph {et~al.}(2004)\citenamefont {Fujita}, \citenamefont {Matsuda}, \citenamefont {Katano},\ and\ \citenamefont {Yamada}}]{fujita2004magnetic}%
  \BibitemOpen
  \bibfield  {author} {\bibinfo {author} {\bibfnamefont {M.}~\bibnamefont {Fujita}}, \bibinfo {author} {\bibfnamefont {M.}~\bibnamefont {Matsuda}}, \bibinfo {author} {\bibfnamefont {S.}~\bibnamefont {Katano}}, \ and\ \bibinfo {author} {\bibfnamefont {K.}~\bibnamefont {Yamada}},\ }\href@noop {} {\bibfield  {journal} {\bibinfo  {journal} {Physical review letters}\ }\textbf {\bibinfo {volume} {93}},\ \bibinfo {pages} {147003} (\bibinfo {year} {2004})}\BibitemShut {NoStop}%
\bibitem [{\citenamefont {Khasanov}\ \emph {et~al.}(2009)\citenamefont {Khasanov}, \citenamefont {Maisuradze}, \citenamefont {Maeter}, \citenamefont {Kwadrin}, \citenamefont {Luetkens}, \citenamefont {Amato}, \citenamefont {Schnelle}, \citenamefont {Rosner}, \citenamefont {Leithe-Jasper},\ and\ \citenamefont {Klauss}}]{khasanov2009superconductivity}%
  \BibitemOpen
  \bibfield  {author} {\bibinfo {author} {\bibfnamefont {R.}~\bibnamefont {Khasanov}}, \bibinfo {author} {\bibfnamefont {A.}~\bibnamefont {Maisuradze}}, \bibinfo {author} {\bibfnamefont {H.}~\bibnamefont {Maeter}}, \bibinfo {author} {\bibfnamefont {A.}~\bibnamefont {Kwadrin}}, \bibinfo {author} {\bibfnamefont {H.}~\bibnamefont {Luetkens}}, \bibinfo {author} {\bibfnamefont {A.}~\bibnamefont {Amato}}, \bibinfo {author} {\bibfnamefont {W.}~\bibnamefont {Schnelle}}, \bibinfo {author} {\bibfnamefont {H.}~\bibnamefont {Rosner}}, \bibinfo {author} {\bibfnamefont {A.}~\bibnamefont {Leithe-Jasper}}, \ and\ \bibinfo {author} {\bibfnamefont {H.-H.}\ \bibnamefont {Klauss}},\ }\href@noop {} {\bibfield  {journal} {\bibinfo  {journal} {Physical review letters}\ }\textbf {\bibinfo {volume} {103}},\ \bibinfo {pages} {067010} (\bibinfo {year} {2009})}\BibitemShut {NoStop}%
\bibitem [{\citenamefont {Sigrist}\ and\ \citenamefont {Rice}(1995)}]{sigrist1995unusual}%
  \BibitemOpen
  \bibfield  {author} {\bibinfo {author} {\bibfnamefont {M.}~\bibnamefont {Sigrist}}\ and\ \bibinfo {author} {\bibfnamefont {T.}~\bibnamefont {Rice}},\ }\href@noop {} {\bibfield  {journal} {\bibinfo  {journal} {Reviews of Modern Physics}\ }\textbf {\bibinfo {volume} {67}},\ \bibinfo {pages} {503} (\bibinfo {year} {1995})}\BibitemShut {NoStop}%
\bibitem [{\citenamefont {Guguchia}\ \emph {et~al.}(2015)\citenamefont {Guguchia}, \citenamefont {Amato}, \citenamefont {Kang}, \citenamefont {Luetkens}, \citenamefont {Biswas}, \citenamefont {Prando}, \citenamefont {von Rohr}, \citenamefont {Bukowski}, \citenamefont {Shengelaya}, \citenamefont {Keller} \emph {et~al.}}]{guguchia2015direct}%
  \BibitemOpen
  \bibfield  {author} {\bibinfo {author} {\bibfnamefont {Z.}~\bibnamefont {Guguchia}}, \bibinfo {author} {\bibfnamefont {A.}~\bibnamefont {Amato}}, \bibinfo {author} {\bibfnamefont {J.}~\bibnamefont {Kang}}, \bibinfo {author} {\bibfnamefont {H.}~\bibnamefont {Luetkens}}, \bibinfo {author} {\bibfnamefont {P.~K.}\ \bibnamefont {Biswas}}, \bibinfo {author} {\bibfnamefont {G.}~\bibnamefont {Prando}}, \bibinfo {author} {\bibfnamefont {F.}~\bibnamefont {von Rohr}}, \bibinfo {author} {\bibfnamefont {Z.}~\bibnamefont {Bukowski}}, \bibinfo {author} {\bibfnamefont {A.}~\bibnamefont {Shengelaya}}, \bibinfo {author} {\bibfnamefont {H.}~\bibnamefont {Keller}},  \emph {et~al.},\ }\href@noop {} {\bibfield  {journal} {\bibinfo  {journal} {Nature communications}\ }\textbf {\bibinfo {volume} {6}},\ \bibinfo {pages} {8863} (\bibinfo {year} {2015})}\BibitemShut {NoStop}%
\bibitem [{\citenamefont {Tafti}\ \emph {et~al.}(2013)\citenamefont {Tafti}, \citenamefont {Juneau-Fecteau}, \citenamefont {Delage}, \citenamefont {Ren{\'e}~de Cotret}, \citenamefont {Reid}, \citenamefont {Wang}, \citenamefont {Luo}, \citenamefont {Chen}, \citenamefont {Doiron-Leyraud},\ and\ \citenamefont {Taillefer}}]{tafti2013sudden}%
  \BibitemOpen
  \bibfield  {author} {\bibinfo {author} {\bibfnamefont {F.}~\bibnamefont {Tafti}}, \bibinfo {author} {\bibfnamefont {A.}~\bibnamefont {Juneau-Fecteau}}, \bibinfo {author} {\bibfnamefont {M.-E.}\ \bibnamefont {Delage}}, \bibinfo {author} {\bibfnamefont {S.}~\bibnamefont {Ren{\'e}~de Cotret}}, \bibinfo {author} {\bibfnamefont {J.-P.}\ \bibnamefont {Reid}}, \bibinfo {author} {\bibfnamefont {A.}~\bibnamefont {Wang}}, \bibinfo {author} {\bibfnamefont {X.}~\bibnamefont {Luo}}, \bibinfo {author} {\bibfnamefont {X.}~\bibnamefont {Chen}}, \bibinfo {author} {\bibfnamefont {N.}~\bibnamefont {Doiron-Leyraud}}, \ and\ \bibinfo {author} {\bibfnamefont {L.}~\bibnamefont {Taillefer}},\ }\href@noop {} {\bibfield  {journal} {\bibinfo  {journal} {Nature Physics}\ }\textbf {\bibinfo {volume} {9}},\ \bibinfo {pages} {349} (\bibinfo {year} {2013})}\BibitemShut {NoStop}%
\bibitem [{\citenamefont {Hicks}\ \emph {et~al.}(2018)\citenamefont {Hicks}, \citenamefont {Ghosh}, \citenamefont {Barber},\ and\ \citenamefont {Klauss}}]{hicks2018piezoelectric}%
  \BibitemOpen
  \bibfield  {author} {\bibinfo {author} {\bibfnamefont {C.~W.}\ \bibnamefont {Hicks}}, \bibinfo {author} {\bibfnamefont {S.}~\bibnamefont {Ghosh}}, \bibinfo {author} {\bibfnamefont {M.~E.}\ \bibnamefont {Barber}}, \ and\ \bibinfo {author} {\bibfnamefont {H.-H.}\ \bibnamefont {Klauss}},\ }in\ \href@noop {} {\emph {\bibinfo {booktitle} {Proceedings of the 14th International Conference on Muon Spin Rotation, Relaxation and Resonance ($\mu$SR2017)}}}\ (\bibinfo {year} {2018})\ p.\ \bibinfo {pages} {011040}\BibitemShut {NoStop}%
\bibitem [{\citenamefont {Guguchia}\ \emph {et~al.}(2020)\citenamefont {Guguchia}, \citenamefont {Das}, \citenamefont {Wang}, \citenamefont {Adachi}, \citenamefont {Kitajima}, \citenamefont {Elender}, \citenamefont {Br\"uckner}, \citenamefont {Ghosh}, \citenamefont {Grinenko}, \citenamefont {Shiroka}, \citenamefont {M\"uller}, \citenamefont {Mudry}, \citenamefont {Baines}, \citenamefont {Bartkowiak}, \citenamefont {Koike}, \citenamefont {Amato}, \citenamefont {Tranquada}, \citenamefont {Klauss}, \citenamefont {Hicks},\ and\ \citenamefont {Luetkens}}]{PhysRevLett.125.097005}%
  \BibitemOpen
  \bibfield  {author} {\bibinfo {author} {\bibfnamefont {Z.}~\bibnamefont {Guguchia}}, \bibinfo {author} {\bibfnamefont {D.}~\bibnamefont {Das}}, \bibinfo {author} {\bibfnamefont {C.~N.}\ \bibnamefont {Wang}}, \bibinfo {author} {\bibfnamefont {T.}~\bibnamefont {Adachi}}, \bibinfo {author} {\bibfnamefont {N.}~\bibnamefont {Kitajima}}, \bibinfo {author} {\bibfnamefont {M.}~\bibnamefont {Elender}}, \bibinfo {author} {\bibfnamefont {F.}~\bibnamefont {Br\"uckner}}, \bibinfo {author} {\bibfnamefont {S.}~\bibnamefont {Ghosh}}, \bibinfo {author} {\bibfnamefont {V.}~\bibnamefont {Grinenko}}, \bibinfo {author} {\bibfnamefont {T.}~\bibnamefont {Shiroka}}, \bibinfo {author} {\bibfnamefont {M.}~\bibnamefont {M\"uller}}, \bibinfo {author} {\bibfnamefont {C.}~\bibnamefont {Mudry}}, \bibinfo {author} {\bibfnamefont {C.}~\bibnamefont {Baines}}, \bibinfo {author} {\bibfnamefont {M.}~\bibnamefont {Bartkowiak}}, \bibinfo {author} {\bibfnamefont {Y.}~\bibnamefont {Koike}}, \bibinfo {author} {\bibfnamefont {A.}~\bibnamefont
  {Amato}}, \bibinfo {author} {\bibfnamefont {J.~M.}\ \bibnamefont {Tranquada}}, \bibinfo {author} {\bibfnamefont {H.-H.}\ \bibnamefont {Klauss}}, \bibinfo {author} {\bibfnamefont {C.~W.}\ \bibnamefont {Hicks}}, \ and\ \bibinfo {author} {\bibfnamefont {H.}~\bibnamefont {Luetkens}},\ }\href {\doibase 10.1103/PhysRevLett.125.097005} {\bibfield  {journal} {\bibinfo  {journal} {Phys. Rev. Lett.}\ }\textbf {\bibinfo {volume} {125}},\ \bibinfo {pages} {097005} (\bibinfo {year} {2020})}\BibitemShut {NoStop}%
\bibitem [{\citenamefont {Grinenko}\ \emph {et~al.}(2021)\citenamefont {Grinenko}, \citenamefont {Ghosh}, \citenamefont {Sarkar}, \citenamefont {Orain}, \citenamefont {Nikitin}, \citenamefont {Elender}, \citenamefont {Das}, \citenamefont {Guguchia}, \citenamefont {Br{\"u}ckner}, \citenamefont {Barber} \emph {et~al.}}]{grinenko2021split}%
  \BibitemOpen
  \bibfield  {author} {\bibinfo {author} {\bibfnamefont {V.}~\bibnamefont {Grinenko}}, \bibinfo {author} {\bibfnamefont {S.}~\bibnamefont {Ghosh}}, \bibinfo {author} {\bibfnamefont {R.}~\bibnamefont {Sarkar}}, \bibinfo {author} {\bibfnamefont {J.-C.}\ \bibnamefont {Orain}}, \bibinfo {author} {\bibfnamefont {A.}~\bibnamefont {Nikitin}}, \bibinfo {author} {\bibfnamefont {M.}~\bibnamefont {Elender}}, \bibinfo {author} {\bibfnamefont {D.}~\bibnamefont {Das}}, \bibinfo {author} {\bibfnamefont {Z.}~\bibnamefont {Guguchia}}, \bibinfo {author} {\bibfnamefont {F.}~\bibnamefont {Br{\"u}ckner}}, \bibinfo {author} {\bibfnamefont {M.~E.}\ \bibnamefont {Barber}},  \emph {et~al.},\ }\href@noop {} {\bibfield  {journal} {\bibinfo  {journal} {Nature Physics}\ }\textbf {\bibinfo {volume} {17}},\ \bibinfo {pages} {748} (\bibinfo {year} {2021})}\BibitemShut {NoStop}%
\bibitem [{\citenamefont {Khasanov}\ \emph {et~al.}(2005)\citenamefont {Khasanov}, \citenamefont {Eshchenko}, \citenamefont {Di~Castro}, \citenamefont {Shengelaya}, \citenamefont {La~Mattina}, \citenamefont {Maisuradze}, \citenamefont {Baines}, \citenamefont {Luetkens}, \citenamefont {Karpinski}, \citenamefont {Kazakov},\ and\ \citenamefont {Keller}}]{Khasanov104504}%
  \BibitemOpen
  \bibfield  {author} {\bibinfo {author} {\bibfnamefont {R.}~\bibnamefont {Khasanov}}, \bibinfo {author} {\bibfnamefont {D.~G.}\ \bibnamefont {Eshchenko}}, \bibinfo {author} {\bibfnamefont {D.}~\bibnamefont {Di~Castro}}, \bibinfo {author} {\bibfnamefont {A.}~\bibnamefont {Shengelaya}}, \bibinfo {author} {\bibfnamefont {F.}~\bibnamefont {La~Mattina}}, \bibinfo {author} {\bibfnamefont {A.}~\bibnamefont {Maisuradze}}, \bibinfo {author} {\bibfnamefont {C.}~\bibnamefont {Baines}}, \bibinfo {author} {\bibfnamefont {H.}~\bibnamefont {Luetkens}}, \bibinfo {author} {\bibfnamefont {J.}~\bibnamefont {Karpinski}}, \bibinfo {author} {\bibfnamefont {S.~M.}\ \bibnamefont {Kazakov}}, \ and\ \bibinfo {author} {\bibfnamefont {H.}~\bibnamefont {Keller}},\ }\href {\doibase 10.1103/PhysRevB.72.104504} {\bibfield  {journal} {\bibinfo  {journal} {Phys. Rev. B}\ }\textbf {\bibinfo {volume} {72}},\ \bibinfo {pages} {104504} (\bibinfo {year} {2005})}\BibitemShut {NoStop}%
\bibitem [{\citenamefont {Suter}\ and\ \citenamefont {Wojek}(2012)}]{Suter69}%
  \BibitemOpen
  \bibfield  {author} {\bibinfo {author} {\bibfnamefont {A.}~\bibnamefont {Suter}}\ and\ \bibinfo {author} {\bibfnamefont {B.}~\bibnamefont {Wojek}},\ }\href {\doibase https://doi.org/10.1016/j.phpro.2012.04.042} {\bibfield  {journal} {\bibinfo  {journal} {Phys. Procedia}\ }\textbf {\bibinfo {volume} {30}},\ \bibinfo {pages} {69} (\bibinfo {year} {2012})}\BibitemShut {NoStop}%
\bibitem [{\citenamefont {Tinkham}(2004)}]{Tinkham2004}%
  \BibitemOpen
  \bibfield  {author} {\bibinfo {author} {\bibfnamefont {M.}~\bibnamefont {Tinkham}},\ }\href {http://app.knovel.com/hotlink/toc/id:kpISE00023/introduction-to-superconductivity} {\emph {\bibinfo {title} {Introduction to superconductivity}}},\ \bibinfo {edition} {2nd}\ ed.\ (\bibinfo  {publisher} {Dover Publications Mineola, N.Y},\ \bibinfo {address} {Mineola, N.Y},\ \bibinfo {year} {2004})\BibitemShut {NoStop}%
\bibitem [{\citenamefont {Carrington}\ and\ \citenamefont {Manzano}(2003)}]{Carrington205}%
  \BibitemOpen
  \bibfield  {author} {\bibinfo {author} {\bibfnamefont {A.}~\bibnamefont {Carrington}}\ and\ \bibinfo {author} {\bibfnamefont {F.}~\bibnamefont {Manzano}},\ }\href {\doibase https://doi.org/10.1016/S0921-4534(02)02319-5} {\bibfield  {journal} {\bibinfo  {journal} {Physica C: Superconductivity}\ }\textbf {\bibinfo {volume} {385}},\ \bibinfo {pages} {205} (\bibinfo {year} {2003})}\BibitemShut {NoStop}%
\end{thebibliography}%
\end{document}